\documentclass[aps,prd,showpacs,nofootinbib,twocolumn]{revtex4}
\usepackage{latexsym}
\usepackage{amsmath,amsfonts}
\usepackage{amsbsy}
\usepackage{mathrsfs}
\usepackage{color}
\usepackage{float}
\usepackage{dsfont}
\usepackage{psfrag}
\usepackage{enumerate}
\usepackage{soul}
\usepackage{url}
\usepackage{hyperref}

\usepackage{amsmath,amssymb,calc,amsfonts}
\usepackage{latexsym}
\usepackage{hyperref}

\usepackage{graphicx,calc,epsfig}

\begin{document}

\author{
Gabriel Crisnejo$^{1,3}$,\thanks{E-mail: gcrisnejo@famaf.unc.edu.ar}
Emanuel Gallo $^{1,3}$\thanks{E-mail: \textcolor{blue}{to complete}}, 
Ezequiel F. Boero$^{1,2}$\thanks{E-mail: ezequiel.boero@unc.edu.ar} and 
Osvaldo M. Moreschi $^{1,3}$\thanks{E-mail: \textcolor{blue}{to complete}}} 
\affiliation{$^{1}$Facultad de Matem\'{a}tica,  Astronom\'\i{}a, F\'\i{}sica y Computaci\'{o}n (FaMAF), 
Universidad Nacional de C\'{o}rdoba, \\ 
Medina Allende S/N, X5000HUA, Córdoba, Argentina.  \\
$^{2}$Instituto de Astronom\'\i{}a Te\'{o}rica y Experimental (IATE), CONICET, \\
Observatorio Astron\'{o}mico de Córdoba,\\
Laprida 854, (X5000BGR) C\'{o}rdoba, Argentina.\\
$^{3}$Instituto de Física Enrique Gaviola, IFEG, CONICET, Medina Allende S/N, X5000HUA, Córdoba, Argentina.
}

\title{Perturbative and numerical approach to plasma strong lensing}

\begin{abstract}
 {Using two different approaches, we study imaging in the strong lens regime taking into account the effects of plasmatic environments on light propagation. First, we extend the use of a perturbative approach that allows us to quickly and analytically calculate the position and shape of the images of a circular source lensed by a galaxy. Such approach will be compared with that obtained from the numerical solution of the lens equations. Secondly, we introduce a 3-dimensional spheroidal model to describe the spacetime associated with the dark matter halo around the lens galaxy and an associated optical metric to incorporate the presence of the plasma medium. The (chromatic) deformation on caustic and critical curves and associated multiplicity of images is also analyzed for particular configurations.}

\end{abstract}


\maketitle

\newcommand{\red}[1]{\textcolor{red}{#1}}
\newcommand{\blue}[1]{\textcolor{blue}{#1}}
\newcommand{\green}[1]{\textcolor{green}{#1}}

\newcommand{\ama}[1]{{\color{ama} #1}}
\newcommand{\vio}[1]{{\color{vio} #1}}
\newcommand{\azul}[1]{{\color{blue} #1}}
\newcommand{\verde}[1]{{\color{verde} #1}}
\newcommand{\marron}[1]{{\color{marron} #1}}
\newcommand{\rojo}[1]{{\color{red} #1}}
\definecolor{azul}{rgb}{0,0,1}
\definecolor{verde}{rgb}{0.,.5,0.4}
\definecolor{marron}{rgb}{0.7,0.2,0.1}
\definecolor{rojo}{rgb}{1,0,0}
\definecolor{vio}{rgb}{0.6,0,0.8}
\definecolor{ama}{rgb}{0,1,1}

\newcommand{\f}[2]{\frac{#1}{#2}}

\section{Introduction}

In general, in the optical geometric limit, electromagnetic radiation propagates along non geodesic curved trajectories when interacting with an in-homogeneous optical medium, as opposed to the geodesic paths in the gravity pure case. While such optical phenomena are well-known in Earthbound laboratories, analogous optical effects have also been observed to occur in astrophysical scenarios across the electromagnetic spectrum. Due to rays passing through an intervening optical medium, a viewer may observe multiple images of distant objects, distorted and apparently shifted from their true locations.  In fact, such cosmic lensing scenarios can display both converging and diverging behaviour depending on the nature of the lenses involved.

Gravitational lensing occurs when light rays passes by a massive object. In this case, the curvature of spacetime behaves as an effective optical medium that depends on the derivatives of components of the metric of the spacetime produced by the lens \citep{Narayan:1996ba, Schneider92, perlick2000ray}. As an example, an isolated point mass generally behaves like a converging lens, acting to magnify a well-aligned distant source. Since all frequencies of radiation are affected equally by spacetime curvature, gravitational lensing effects are achromatic. At present, gravitational lensing is known on many scales, ranging from micro-arcsecond images arising from individual stellar lenses \citep{Mao:2012za} to galaxies \citep{Koopmans:2009av} and entire galaxy clusters \citep{Clowe:2006eq}.

In contrast to gravitation, the ionized matter of the interstellar medium (ISM) may also act as an optical medium to affect our view of distant objects, albeit an optical medium with vastly different properties than gravitational fields. The plasma component of the turbulent and clumpy ISM may obscure background radio sources, causing demagnification analogous to a diverging lens \citep{Clegg:1997ya}. Due to the frequency dependence of the plasma dispersion relation, the lensing effect of plasma lenses are chromatic, strongly altering the paths of radio waves\cite{Bisnovatyi-Kogan:2010flt,Crisnejo:2018uyn,Crisnejo:2018ppm,Crisnejo:2019xtp,Crisnejo:2019ril}. It is believed that such plasma structures are responsible for extreme scattering events (ESEs)\citep{Tuntsov_2016ApJ, Bannister_2016Sci}, in which the flux density of radio-loud background sources ($ 1$ GHz) are observed to dim by large amounts (exceeding $50\%$) \citep{Stanimirovi_2018}. The ESE phenomena may also be closely related to pulsar scintillation, which has been the motivation behind many exotic plasma distributions beyond simple spherical symmetry such as filaments \citep{Suresh_2019ApJ, Rogers_2020MNRAS} and sheets of ionized matter \citep{Pen:2013njl, Simard_2018MNRAS}. In fact, plasma lenses do not display exclusively diverging optical behaviour, since they can also be under-densities in an over-dense background. Such 'pits' then behave like converging chromatic lenses \citep{Pen:2011dz}, however recent observations seem to exclude this possibility at least for a subset of observed ESEs \citep{Bannister_2016Sci}. One of the most striking observations of plasma lensing has occurred due to the observation of a black widow pulsar through the ablated wind of its companion \citep{Main_2018Natur, Li_2019MNRAS}. Plasma lenses have also been suggested to affect the propagation of fast radio bursts \cite{2020ApJ...889..158E,Er:2022lad}. Therefore, plasma lenses are highly relevant to topics on the leading edge of astrophysics. However, many aspects of these phenomena remain mysterious, in particular the small physical extent of spherically symmetric models suggest extremely large central densities \citep{Stanimirovi_2018}. Therefore, further study of these enigmatic structures is necessary.

In this work we are interested in the study of lens effects produced by the presence of plasma in strong lensing situations, that is, when multiple source images are produced and they are highly distorted. Some studies on strong gravitational lensing+plasma phenomena have been previously analyzed by various authors; for example to analyze how the presence of plasma atmospheres in neutron stars affects the luminosity curves\cite{Rogers_2015,Rogers:2016xcc,Battye:2021xvt,Briozzo:2022yzi}, or the shadow of black holes \cite{Perlick:2017fio, Perlick:2021aok, Kimpson:2019mji,Huang:2018rfn,Zhang:2022osx,Badia:2021kpk,Briozzo:2022mgg} (strong lensing+strong field effects); or even to the study of arc lensing formation (through ray tracing numerical integration) when the acting lenses are formed by galaxies \cite{Er:2013efa}, or in a situations of microlensing \cite{Tsupko:2019axo,Sun:2022ujt}  (strong lensing + weak field). Note that as in \cite{Er:2013efa} the ray tracing integration has been carried out numerically, it is difficult in many situations to make a more detailed study of how the different parameters that describe the plasma affect the formation of images and their multiplicity. It is our intention to contribute with a more detailed study on the effect of plasma in galaxies on images from distant sources, for which we will present a series of contributions: a) we analyze more generic plasma models than those usually studied that are spherically symmetric, b ) we present a perturbative formulation of the solutions to the lens equation, when  in addition to the gravitational field,  the presence of plasma is taken into account, which will allow us to give analytical formulas to describe the images c) a lens given described by a 3-dimensional spheroidal model and the study of light rays in the strong lens regime. 

{This paper is organized as follows. In Sec.\eqref{sec:Basic_equations} we present the basic equations and an extension of a perturbative method to solve the lens equations originally introduced by Alard\cite{Alard:2007ya} for the gravitational pure case. We also present a general formula for an iterative solution of the perturbative equations. In \eqref{galaxy_modelling} we present the models for the gravitational potentials and plasma profiles, and their effect on the lensed images. In Sec.\eqref{sec4} we carry out a comparison between the analytical  and the numerical solutions and the improvement that successive iterations can produce. In Sec.\eqref{caustic-critical} we describe the critical and caustic curves and the multiplicity of the images for the considered models. In Sec.\eqref{3D-model} we change our approach, introducing a full 4-dimensional metric describing a 3D spheroidal model for the dark matter halo of the galaxy that act as a lens. Different optical scalars, critical and caustic curves and images of the lensed sources are also studied.}

Throughout this paper we assume flat $\Lambda$CDM cosmology with $\Omega_m = 0.315$,  and $H_0 = 67.4\frac{\text{km} \, \text{s}^{-1}}{\text{Mpc}}$, based on the observations of the Planck collaboration\cite{2020A&A...641A...6P}. In addition, we consider a lens and a source with redshift $z_l=0.04$ and $z_s=0.1$, respectively.

\section{Basic equations}
\label{sec:Basic_equations}

In the thin lens approximation, the lens equation relating the positions of the source to those of the images through the angle of deflection and can be written as follows,
\begin{equation}\label{eq:lens+equation}
    \vec{\beta} = \vec{\theta}-\alpha,
\end{equation}
where $\vec{\beta}$, $\vec{\theta}$ denote the angular position of the source and of the image, respectively, and $\alpha$ is related to the deflection angle $\hat\alpha$ by $\alpha=\frac{D_{ls}}{ D_s}\hat\alpha$. In addition, $D_l$, $D_s$ and $D_{ls}$ indicate the angular diameter distance of the observer to the lens, the observer to the source and the lens to the source.

Let us consider as a model for the lens a static and asymptotically flat spacetime, with coordinates $\{x^0,x^i\}$, with $x^i$, $i=1..3$ being spacelike coordinates. On the other hand, we will also consider in this section that the spatial components of the energy-momentum tensor are negligible, so the gravitational lens (not counting the plasma) will be completely described by its matter distribution. Under this assumption the lens equation can be rewritten in terms of the lens potential $\psi_{\text{grav}}$ as follows,
\begin{equation}\label{eq:lente2}
    \vec{\beta} = \vec{\theta}-\nabla_{\vec{\theta}}\psi_{\text{grav}}(\vec{\theta}),
\end{equation}
which is related to the the deflection angle $\alpha$ through $\alpha(\vec{\theta})=\nabla_{\vec{\theta}}\psi_{\text{grav}}(\vec{\theta})$, where $\nabla_{\vec{\theta}}$ is the gradient with respect to angular position $\vec{\theta}$ in the lens plane. We refer $\psi_{\text{grav}}$ as the effective lensing potential and it is related to the Newtonian potential $\Phi$ as follows,
\begin{equation}
    \psi_{\text{grav}}(\vec{\theta})=\frac{D_{ls}}{D_l D_s}\frac{2}{c^2}\int \Phi(D_l\vec{\theta},x)dx,
\end{equation}
being $x$ the line of sight distance between the observer and the source

Since we are interested in the average plasma medium surrounding galaxies, for our purposes, it will suffice to consider it as a cold, non-magnetized plasma. Neglecting the birefringence effects of galactic magnetic fields is based on the assumption that they are generally expected to be of very low intensity (of the order of a few $\mu G$ \cite{Beck:2013bxa}), and therefore their effects on image formation will be negligible. However, they could be considered if one wishes to study the polarization of light due to the Faraday rotation effect\cite{2020MNRAS.493.1736R}.  In turn, magnetic fields must be taken into account in situations of plasma lensing in the vicinity of compact objects. In that situations, the plasma lensing effect can be useful in constraining the value of magnetic fields\cite{Li:2018ssw}.

As the observation frequencies will generally be higher than the associated plasma frequencies, the ISM can be considered an electromagnetic continuous medium with an associated refractive index \cite{2020ApJ...889..158E,bittencourt2013fundamentals,draine2010physics,Donner:2019hlp}.
\begin{equation}
    n^2(x)=1-\frac{\omega_e^2(x^i)}{\omega^2(x^i)},
\end{equation}
where $\omega(x^i)$ is the frequency of the light ray at $x^i$ and the electron plasma frequency $\omega_e(x^i)$ is related to the electron number density $n_e(x^i)$ as follows,
\begin{equation}
    \omega^2_e(x^i) = \frac{4\pi e^2}{m_e^2}n_e(x^i),
\end{equation}
where $e$ and $m_e$ are the electron charge and the electron mass, respectively.

In analogy with the gravitational lensing theory, the cold non-magnetized plasma effect on the light rays can be codified in a frequency-dependent effective potential \cite{Tuntsov_2016ApJ},
\begin{equation}\label{phi_plasma_basic_eq}
    \Phi(x^i) \approx \frac{c^2 \omega^2_e(x^i)}{4\omega^2(x^i)}.
\end{equation}

In \ref{phi_plasma_basic_eq} we use the large observational frequency limit, $\omega \gg \omega_e$, which is suitable in general astrophysical situations. In turn, on its way from the source to the observer, the light will experience a gravitational redshift due to the presence of the lens and a cosmological redshift, the former being negligible with respect to the latter because the lens-observer distance is large enough. For this reason we will only take into account the effect of the cosmological redshift so that the frequency of the photons in the position of the lens will be $(1+z_l)\omega$, where $z_l$ is the cosmological redshift of the lens and $\omega$ is the observational measured angular frequency.

Considering a light ray propagating 
through the plasma in a $\hat{x}$ 
direction, the electron column density
$N_e$ is given by the following
expression usually known as dispersion
measure,

\begin{equation}\label{N_e}
    N_e(\vec{\theta}) = \int n_e(x^i(x))dx,
\end{equation}
which can be estimated from time delay measurements.
Finally, the effective lensing potential in terms of the observational frequency $\omega$, the effective lensing potential is given by \cite{Tuntsov_2016ApJ},
\begin{equation}\label{psi_plasma}
    \psi_{\text{plasma}}(\vec{\theta},\omega)=\frac{D_{ls}}{D_s D_l}\frac{2\pi c^2}{\omega^2(1+z_l)^2}r_e N_e(\vec{\theta}),
\end{equation}
where $ r_e = \frac{e^2}{m_e c^2} $ is the classical electron radius and the observation frequency $\nu$ measured in Hz is related to $\omega$ by $\nu=\frac{\omega}{2\pi}$. 
Therefore, the total deflection angle will be given by $\alpha(\vec{\theta})=\nabla_{\vec{\theta}}(\psi_{\text{grav}}(\vec{\theta})+\psi_{\text{plasma}}(\vec{\theta},\omega))$. Note that the total deflection angle in this approximation is curl-free. Hence, the associated optical scalars can be described by the shear and convergence as in standard gravitational lensing theory. It is also worth mentioning that, in contrast to the convergent effect on light rays produced by gravitational fields, plasma lensing can produce a divergent effect if there is an over-density of electronic charge with respect to the ISM. In contrast, plasma lenses that are under-dense compared to the surrounding ISM will produce converging lenses and magnify background sources.  

It is important to note that in this work we will not consider the plasma-gravity interaction which in the situations under consideration is negligible compared to the pure gravity contribution as well as the pure plasma one.

\subsection{Perturbative solution of the lens equation}\label{subsec:Pert_sol_approch}

In the following we will briefly review a perturbative solution of the lens equation in the strong gravitational lensing regime that is often accurate for describing gravitational arcs as well as multiple images. This method introduced by Alard (see \cite{Alard:2007ya}) starts from an exact solution of the gravitational lens equation for a point source aligned with the line of sight for a spherically symmetric lens, resulting in a circular image of the source generally known as Einstein ring. 

Let us consider a circular source with (angular) radius $\delta\beta_s$ centered at $(\delta\beta_{10},\delta\beta_{20})$ in the source plane. Explicitly,
\begin{equation}
\delta\vec{\beta} = 
\begin{pmatrix}
\delta\beta_1 \\
\delta\beta_2  \\
\end{pmatrix}
=
\begin{pmatrix}
\delta\beta_{10}+\delta\beta_s \cos\phi_s \\
\delta\beta_{20}+\delta\beta_s \sin\phi_s  \\
\end{pmatrix}, \ \ \text{with} \ \ 0\leq\phi_s\leq 2\pi.
\end{equation}
On the other hand, each point ot the boundary of the source will have an image position $\vec{\theta}$ that can be written as,
\begin{equation}
\vec{\theta} = 
\begin{pmatrix}
\theta_1 \\
\theta_2  \\
\end{pmatrix}
=
\begin{pmatrix}
\theta \cos\phi \\
\theta\sin\phi  \\
\end{pmatrix}, \ \ \text{with} \ \ 0\leq\phi\leq 2\pi,
\end{equation}
with $\theta=|\vec{\theta}|$ depending on $\phi$.

Therefore, in a gravitational lensing system where the effect of the plasma surrounding the lens is also taken into account, the lensing equation is given as follows,
\begin{eqnarray}
    \delta\beta_{10}+\delta\beta_s \cos\phi_s&=&\theta\cos\phi-\cos\phi\frac{\partial\psi}{\partial\theta}+\frac{\sin\phi}{\theta}\frac{\partial\psi}{\partial\phi},\\
    \delta\beta_{20}+\delta\beta_s \sin\phi_s&=&\theta\sin\phi-\sin\phi\frac{\partial\psi}{\partial\theta}-\frac{\cos\phi}{\theta}\frac{\partial\psi}{\partial\phi}.
\end{eqnarray}
where
\begin{equation}
    \psi=\psi(\theta,\phi) = \psi_{\text{grav}}(\theta,\phi)+\psi_{\text{plasma}}(\theta,\phi).
\end{equation}
In particular, these equations imply the following \citep{Habara11}, 
\begin{equation}\label{eq:lensexact}
    \theta=\delta\beta_{10}\cos\phi+\delta\beta_{20}\sin\phi+\frac{\partial\psi}{\partial\theta}\pm\sqrt{\Delta_{\text{exact}}},
\end{equation}
with 
\begin{equation}\label{Deltaexact}
    \Delta_{\text{exact}}=\delta\beta_s^2-\left(\frac{1}{\theta}\frac{\partial\psi}{\partial\phi}-\delta\beta_{10}\sin\phi+\delta\beta_{20}\cos\phi\right)^2.
\end{equation}
Note that eq.\eqref{eq:lensexact} is an exact and implicit equation for $\theta$.

We will now review the perturbative method introduced by Alard to solve the lens equation and we will present a new iterative expression that improves its first order approximation. It should be noted that this method was presented in the case of pure gravity and here we are extending it to the case with plasma.

Let us start by considering a spherically symmetric lens characterized by a lens potential $\psi_0(\theta)$ around the line of sight. This potential can have the contribution of both gravitational field and the plasmatic medium.  Let us consider a point source located along the line of sight at $\vec{\beta} = 0$. In this situation the lens equation reduces to,
\begin{equation}\label{lens-zero}
    \theta-\frac{\partial\psi_0({\theta})}{\partial{\theta}}=0.
\end{equation}

We denote its solutions as ${\theta}_p$. In the pure gravity case, these solutions are known as the Einstein radius and we will denote them as $\theta_E$ or $|\vec{\theta}_E|$. As we will see, the solution ${\theta}_p$ will also have a ring shape in the image plane, but we will reserve the term Einstein ring for the pure gravity case.

On the other hand, we will work with mass and plasma profiles that are commonly used in astrophysics  and that present these two characteristics:  a) there is a solution to equation \eqref{lens-zero} and b) they present a unique circular solution in the plane of the images (Einstein ring in pure gravity case). Other plasma profiles, different from those used in this work, allow us to obtain several of these circular solutions  to equation \eqref{lens-zero}, such as polynomial plasma models \cite{Tsupko:2019axo}. Although the perturbative method presented in this section does not present any impediment to treat these cases, we will leave its study for later works.

Let us now consider a small deviation in the position of the source as well as in the circular symmetry of the lens potential, that is, we will introduce a small ellipticity in the lens potential,
\begin{equation}
\begin{aligned}\label{eq:pertprinci}
    \vec{\beta}=\,&\delta\vec{\beta}, \\
    \psi(\vec{\theta})=\,&\psi_0(|\vec{\theta}|)+\delta\psi(\vec{\theta}).
\end{aligned}    
\end{equation}
We will also assume that these small deviations in the circular symmetry and in the position of the source will imply a small deviation of the background solution, which presents a circular shape in the plane of the images. Then introducing the following decomposition,
\begin{equation}\label{eq:pertthetap}
    \vec{\theta}=\vec{\theta}^p_E + \delta\vec{\theta},
\end{equation}
the lens equation can be rewritten as,
\begin{equation}
    \delta\vec{\beta} = \vec{\theta}^p_E + \delta\vec{\theta} - \nabla_{\vec{\theta}}[\psi_0(|\vec{\theta}|)+\delta\psi(\vec{\theta})]_{\vec{\theta}=\vec{\theta}_p + \delta\vec{\theta}}.
\end{equation}
Even more, we assume the three quantities $\delta\vec{\beta}$, $\delta\psi(\vec{\theta})$ and $\delta\vec{\theta}$ are of the same order of magnitude. To this order the perturbed lens equation is written as follows,
\begin{equation}
    \delta\vec{\beta} = \delta\vec{\theta}-[(\delta\vec{\theta}\cdot\nabla_{\vec{\theta}})\nabla_{\vec{\theta}}\psi_0(|\vec{\theta}|)+\nabla_{\vec{\theta}}\delta\psi(\vec{\theta})]|_{\vec{\theta}=\vec{\theta}_p}.
\end{equation}

Let us consider a circular source located at $(\delta\beta_{10},\delta\beta_{20})$ with radius $\delta\beta_s$. Explicitly,
\begin{equation}
\delta\vec{\beta} = 
\begin{pmatrix}
\delta\beta_1 \\
\delta\beta_2  \\
\end{pmatrix}
=
\begin{pmatrix}
\delta\beta_{10}+\delta\beta_s \cos\phi_s \\
\delta\beta_{20}+\delta\beta_s \sin\phi_s  \\
\end{pmatrix}, \ \ \text{with} \ \ 0\leq\phi_s\leq 2\pi.
\end{equation}
On the other hand, the image position can be written as,
\begin{equation}
\vec{\theta} = 
\begin{pmatrix}
\theta_1 \\
\theta_2  \\
\end{pmatrix}
=
\begin{pmatrix}
\left({\theta}_p+\delta{\theta}\right) \cos\phi \\
\left({\theta}_p+\delta{\theta}\right) \sin\phi  \\
\end{pmatrix}, \ \ \text{with} \ \ 0\leq\phi\leq 2\pi,
\end{equation}
with $\theta^p_E=|\vec{\theta}^p_E|$.

At leading order the two components of the lens equations are then given by,
\begin{equation}
\begin{aligned}
    &\delta\beta_{10}+\delta\beta_s\cos\phi_s= \\ &\left[\delta{\theta}\cos\phi\left(1-\frac{\partial^2\psi_0}{\partial|\vec{\theta}|^2}\right)-\cos\phi\frac{\partial\delta\psi}{\partial|\vec{\theta}|}+\frac{\sin\phi}{|\vec{\theta}|}\frac{\partial\delta\psi}{\partial\phi}\right]\bigg|_{|\vec{\theta}|={\theta}_p}, \\
    &\delta\beta_{20}+\delta\beta_s\sin\phi_s= \\ &\left[\delta{\theta}\sin\phi\left(1-\frac{\partial^2\psi_0}{\partial|\vec{\theta}|^2}\right)-\sin\phi\frac{\partial\delta\psi}{\partial|\vec{\theta}|}-\frac{\cos\phi}{|\vec{\theta}|}\frac{\partial\delta\psi}{\partial\phi}\right]\bigg|_{|\vec{\theta}|={\theta}_p},
\end{aligned}
\end{equation}
and combining these equations we finally obtain,
\begin{equation}\label{ec-lente-primer-orden}
    \delta{\theta}=\frac{1}{1-\frac{\partial^2\psi_0}{\partial|\vec{\theta}|^2}}\left[\frac{\partial\delta\psi}{\partial|\vec{\theta}|}+\delta\beta_{10}\cos\phi+\delta\beta_{20}\sin\phi\pm\sqrt{\Delta}\right]\bigg|_{|\vec{\theta}|={\theta}_p},
\end{equation}
where
\begin{equation}\label{Delta}
    \Delta=\delta\beta_s^2-\left(\frac{1}{|\vec{\theta}|}\frac{\partial\delta\psi}{\partial\phi}-\delta\beta_{10}\sin\phi+\delta\beta_{20}\cos\phi\right)^2.
\end{equation}
From \eqref{Delta} we can see that the region on the image plane where the formation of images are allowed is characterized by the condition
\begin{equation}\label{image-condition}
    \Delta\left(|\vec{\theta}|={\theta}_p,\phi\right)\geq0.   
\end{equation}

\subsection{N-iteration's formula}
\label{N-iteraciones}

In order to construct the first approximation to the image position we started with the image at order zero placed in $\vec{\theta}=\vec{\theta}^p_E$ and we find the corrected images placed at $\vec{\theta}^p_1=\vec{\theta}^p_E+\delta\vec{\theta}^p_1$. This method can be easily extended
to higher orders. Assuming that the images at the $i-1$-th iteration is known, we perturb the potentials around these positions and computed new corrections. That is, knowing $\vec{\theta}^p_{i-1}$, we can construct a new correction given by $\vec{\theta}^p_i=(|\vec{\theta}^p_{i-1}|+\delta{\theta}^p_i)(\cos\phi,\sin\phi)$. After repetition of the same procedure as before, we find 
\begin{equation}\label{eq:corriter}
    \delta\theta^p_i=\frac{\frac{\partial\psi_0(\theta)}{\partial|\theta|}-\theta+c}{1-\frac{\partial^2\psi_0}{\partial|\theta|^2}}\bigg|_{\vec{\theta}=\vec{\theta}^p_{i-1}}
\end{equation}
where
\begin{equation}
    c=\left[\frac{\partial\delta\psi}{\partial|\vec{\theta}|}+\delta\beta_{10}\cos\phi+\delta\beta_{20}\sin\phi\pm\sqrt{\Delta}\right]\bigg|_{|\vec{\theta}|={\theta}_{i-1}^p}.
\end{equation}
Eq.\eqref{eq:corriter} determines the image position in an iterative way. As we will shown in the following, in most cases a first iteration is sufficiently to compute in an analytical way the (approximate) shapes and location of the images.
Despite the simplicity of \eqref{eq:corriter} we have not knowledge of a previous presentation of this formula in literature.

\section{Galaxy modelling and image formation}\label{galaxy_modelling}

In order to study the effect of plasma on image formation as well as its influence on the structure of caustic and critical curves we need to specify first the mass density profile of the lens or alternatively its lensing potential, and second its electron density profile. In the first part of this work we will only consider galaxy lenses with an 
 gravitational elliptical lensing potential modelled by the singular isothermal elliptical model (SIE) which is widely used to model dark matter halos in galaxies both in the theory of gravitational lensing and in studies of stellar dynamics. In section \ref{3D-model} we will perform a detailed analysis of lensing effect produced by a specific 3D spheroidal model which is an exact solution of the Einstein equations. 

SIE profile is characterized by the following lens potential,
\begin{equation}
    \psi_\text{grav}(\vec{\theta})= {\theta}_E {\theta}\sqrt{1-\eta\cos 2\phi}
\end{equation}
where $\eta$ is the ellipticity, $\theta\equiv|\vec{\theta}|$ and $\theta_E\equiv|\vec{\theta}_E|$ is the Einstein ring which is given in terms of the velocity dispersion $\sigma_c$ as follows, 
\begin{equation}
    \theta_E = 4\pi\frac{\sigma_c^2}{c^2} \frac{D_{ls}}{D_{s}}.
\end{equation}

Regarding the electron density profile models around galaxies, we will consider different continuous distributions. Some of these models were introduced in the literature to fit data from other galaxies \cite{gutierrez2010galaxy, Davies_2021, Chy_y_2018} or our own galaxy \cite{Cordes:2002wz, Yao_2017, Yamasaki:2019htx, Price:2021gzo}. In the latter case, these models (with different levels of sophistication) were constructed from the dispersion measure associated to pulsars and also take into account the contribution of the Magellanic Clouds and the intergalactic medium. These models are very useful to analyze distances to pulsars in our own galaxy. Moreover, in \cite{Chy_y_2018}, the LOFAR Multifrequency Snapshot Sky Survey (MSSS) was used to investigate the radio continuum spectra for a large sample of nearby star-forming galaxies using some of the models discussed here.

In general, the observation frequency will be larger than the plasma frequency. Therefore, the plasma environment will produce a small difference in the position and shape of the images with respect to the high-frequency optical limit (pure gravity case). However, as we will see in the next sections, for particular electron density profiles, observation frequencies and for some given orientation of the observer, the multiplicity of images can change. We refer to \cite{Gupta:1999nf,Pushkarev:2013zqa,Wang:2022kkk} for observational works on multiple imaging in plasma lensing.

We will start with a spherically symmetric electron distribution with an exponential decay and then we will consider other less restrictive profiles.

\begin{figure}
	\includegraphics[width=\columnwidth]{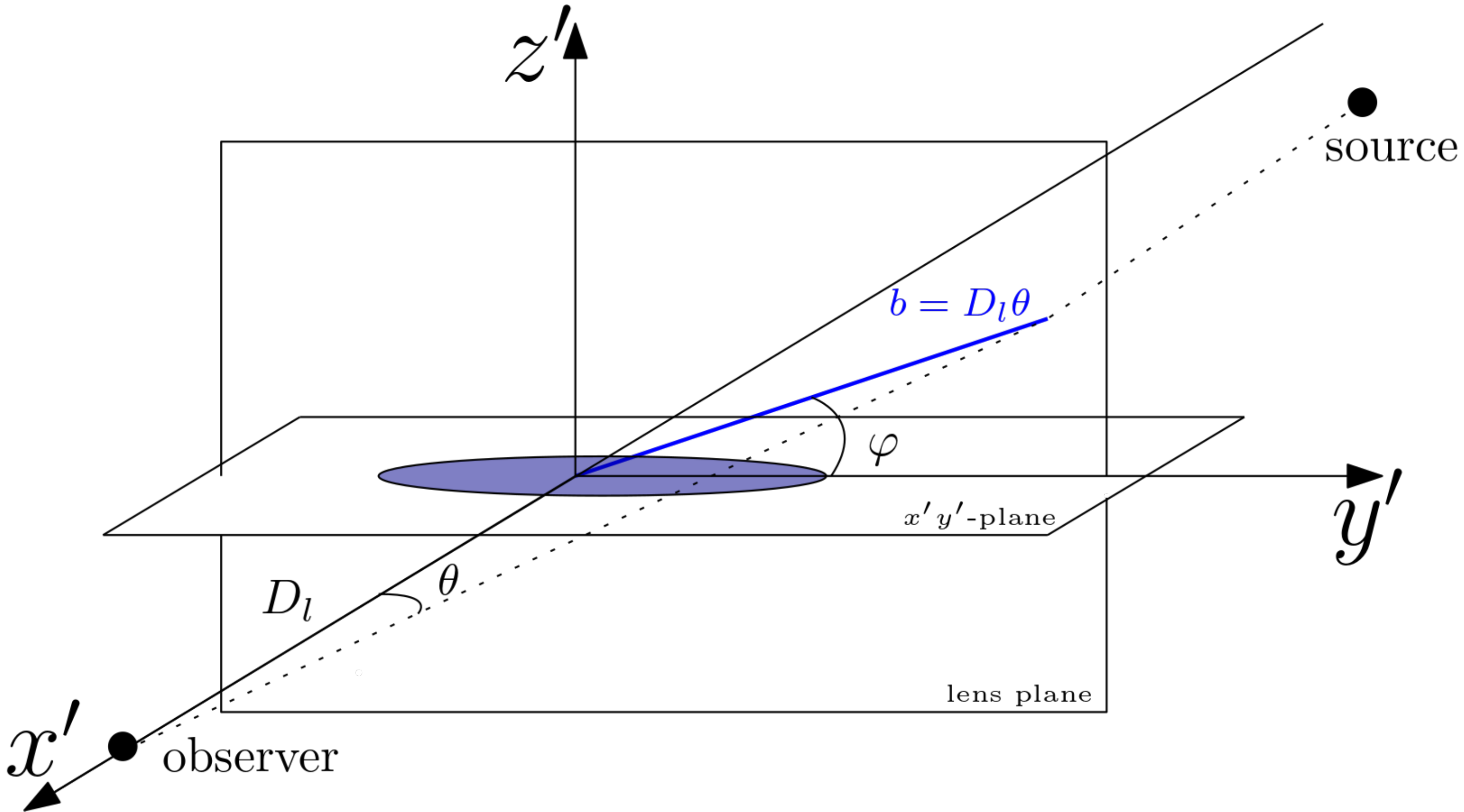}
    \caption{Coordinate system used centered at the position of the lens. The observer is placed at an angular distance $D_l$ from the lens. A light ray coming from the source intersect the lens plane at an angle $\phi$ and at a distance (impact parameter) $b=D_l\phi$ from the center of the lens.}
    \label{coord}
\end{figure}

First of all, we need to specify the coordinate system that we will use. As we can see en Fig. \ref{coord} we have chosen the $x^\prime$-axis in such a way that it coincides with line of sight while the plane of the lens coincides with the $y^\prime z^\prime$ one. In addition, we have defined in the same plane the angle $\varphi$ as shown and a radial cylindrical coordinates $r_c$ is defined in the $x^\prime y^\prime$-plane (not shown). We reserve $r$ as a spherical radial coordinate. Then it is straightforward to check the following relationships,
\begin{eqnarray}\label{coord-strong}
    &y^\prime=b\cos\varphi=D_l \theta \cos\varphi, \\ \label{coord-strong-1}
    &z^\prime=b\sin\varphi=D_l \theta \sin\varphi, \\
    &r_c=\sqrt{x^{\prime 2}+(D_l\theta\cos\varphi)^2}, \\
    &r=\sqrt{x^{\prime 2}+(D_l\theta)^2}.
\end{eqnarray}
In the next subsections we describe the plasma models we will use in this work.

\subsection{Spherically symmetric plasma model with exponential decay}

As a first model we will consider a spherically symmetric electron density with exponential decay given by the following expression,
\begin{equation}\label{exponential_density}
    n_e(r)=n_0e^{-r/r_p}.
\end{equation}
The effect of this kind of plasma profiles on image formation has already been numerically analyzed in the past by Er and Mao in \cite{Er:2013efa}. Here we will develop a perturbative analysis. In turn, this kind of model will serve as a seed to obtain the dispersion measure in more generic models that do not respect spherical symmetry and that will be discussed in the next subsections. This model is inspired by observational fits proposed in the past to study the distribution of ions in H II regions of several galaxies. In \cite{gutierrez2010galaxy}, based on observations of the galaxy M51 (a galaxy facing the line of sight), values of the electron density $n_0 = 10\text{cm}^{-3}$, and $1$kpc for the characteristic radius $r_p$ were estimated. Although in such a galaxy $r$ fulfills the role of galactocentric radius, measured in the direction of the plane that contains the galaxy, in equation \eqref{exponential_density} $r$ is assumed to be a spherical coordinate (it will be relaxed in the next subsections).

\begin{figure}
	\includegraphics[width=\columnwidth]{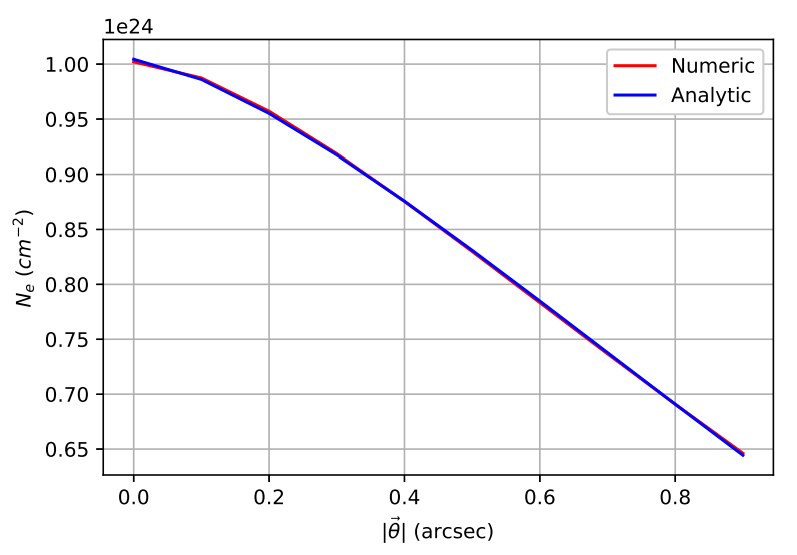}
    \caption{The error on the fitting is less than 0.25$\%$ in this range for a spherically symmetric plasma model with exponential decay and parameters $n_0=60$cm$^{-3}$, $r_p=1$kpc.}
    \label{fig:Ne_fitting}
\end{figure}

\begin{figure}
\centering
\includegraphics[width=\columnwidth]{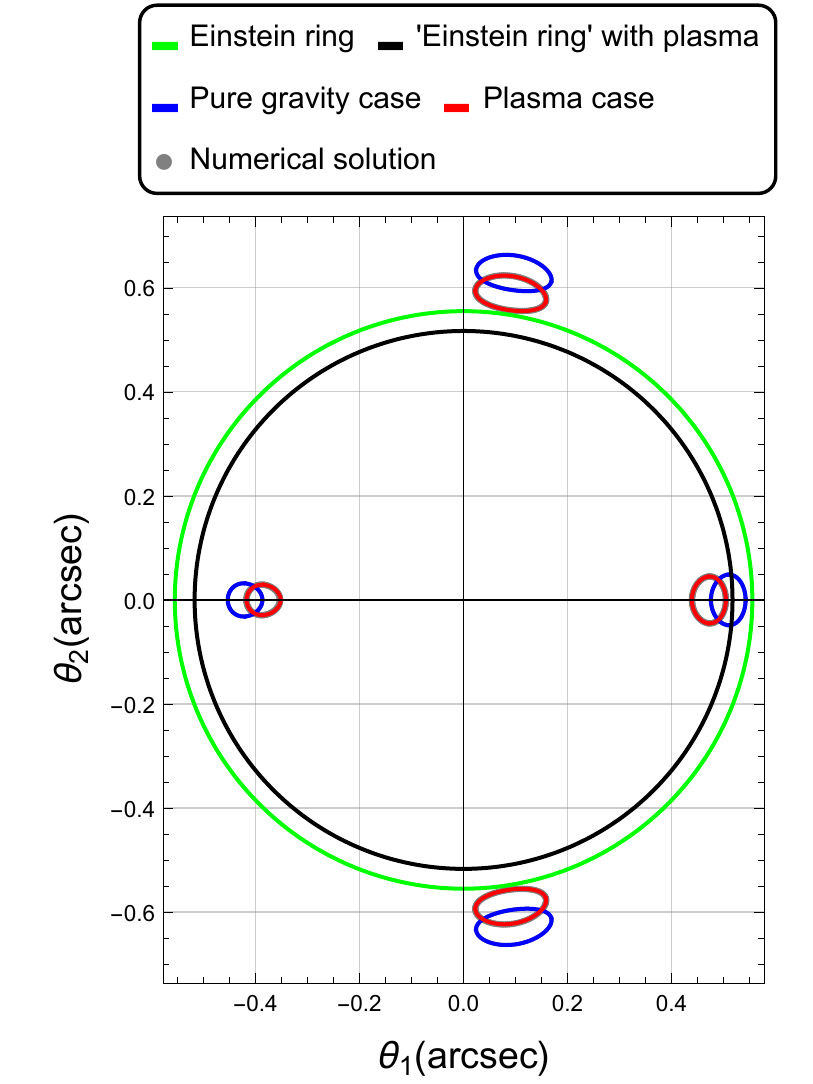}
\caption{SIE model with spherical plasma and exponential decay. $n_0=60$cm$^{-3}$, $r_p=1$kpc, $\sigma_c=180$km/s, $\eta=0$.$3$ at an observation frequency $\nu=80$Mhz. The angular radius of the Einstein ring for this configuration is $\theta_E=0$.$555$arcsec (green line) while $\theta_E^p=0$.$517$arcsec (black line). The source parameters are (radius and position): $\delta\beta_s=0$.$06\theta_E$, $\delta\beta_{10}=0$.$08\theta_E$, $\delta\beta_{20}=0$.$0$.}
\label{hhh}
\end{figure}

Because of the integral \eqref{N_e} cannot be solved analytically for this specific plasma model, we cannot obtain an analytical expression for the electron column density $N_e$. Far from being a limitation of this model, since such integral can always be solved numerically, we chose to fit a function in a suitable range in order to obtain an analytical expression for the solution of the lens equation. Then, we approximate the electron column density $N_e$ as follows,
\begin{equation}\label{Ne_approx}
    N_e(\theta)\approx A n_0 r_p e^{-(\theta/B\theta_0)^C},
\end{equation}
where $\theta_0=r_p/D_l$, and $A$, $B$ and $C$ are dimensionless parameters that we obtain from the fitting. Note that the value of $A, B$ and $C$ will depend on the particular choice of $r_p$. In Fig \ref{fig:Ne_fitting} we graphically show as an example the fitting implemented in equation \eqref{Ne_approx} for this kind of plasma with parameters $n_0=60$cm$^{-3}$ and $r_p=1$kpc. We see that in the range under consideration,  which corresponds to the range where we will have images (see Fig. \ref{hhh}), the implemented fitting is quite adequate with an error of less than 0.25$\%$, while the fitting parameters remain as follows: $A=2$.$003\pm 0$.$002$, $B=1$.$55\pm 0$.$01$, $C=1$.$47\pm 0$.$01$.

The plasma lensing potential for this model reads,
\begin{equation}\label{spherical_plasma}
    \psi_{\text{plasma}}(\theta,\omega)=\frac{D_{ls}}{D_s D_l}\frac{2\pi c^2}{\omega^2(1+z_l)^2} A r_e n_0 r_p e^{-\left(\frac{\theta}{B\theta_0}\right)^C}.
\end{equation}
With the intention to simplify the expressions in the following Sections we rewrite \eqref{spherical_plasma} as,
\begin{equation}\label{spherical-0}
    \psi_{\text{plasma}}(\theta,\omega)=\psi_\omega^2 e^{-\left(\frac{\theta}{B\theta_0}\right)^C},
\end{equation}
where
\begin{equation}\label{spherical-1}
    \psi_\omega^2=\frac{D_{ls}}{D_s D_l}\frac{2\pi c^2}{\omega^2(1+z_l)^2} A r_e n_0 r_p.
\end{equation}
In order to apply the perturbative approach, we take into account the circular symmetry of the plasma potential (in contrast to the SIE gravitational potential), and therefore it will be convenient to consider the perturbation on the total projected lensing potential $\psi_{\text{tot}}(\vec{\theta})$ (i.e. gravitational potential + plasma potential) only due to the deviation from circularity of the gravitational potential. That is, using the notation of the previous section we set 

\begin{widetext}
\begin{eqnarray}
 \psi_{\text{tot}}(\vec{\theta})&=&\psi_{\text{grav}}(\vec{\theta})+\psi_{\text{plasma}}(\vec{\theta},\omega),\\
    \psi_0(\vec{\theta})&=& \psi_{\text{grav}}(\vec{\theta})|_{\eta=0}+\psi_{\text{plasma}}(\vec{\theta},\omega)=\theta_E \;\theta+\psi_{\text{plasma}}(\vec{\theta},\omega)=\theta_E \;\theta+\psi_\omega^2 e^{-\left(\frac{\theta}{B\theta_0}\right)^C},\\
    \delta{\psi}(\vec{\theta})&=&\psi_{\text{tot}}(\vec{\theta})-\psi_0(\theta)= \theta_E \;\theta(\sqrt{1-\eta\cos 2\phi}-1).
\end{eqnarray}

Finally, from \eqref{ec-lente-primer-orden}, the perturbative solution of the lens equation can be expressed as,
\begin{equation}\label{SIS_solution}
    \delta\theta^\pm=\frac{|\vec{\theta}_E|\left(\sqrt{1-\eta\cos2\phi}-1\right)+\delta\beta_{10}\cos\phi+\delta\beta_{20}\sin\phi \pm \sqrt{\Delta}}{1-\frac{\psi_\omega^2}{\theta_p^2}C\zeta e^{-\zeta}(1+C(\zeta-1))},
\end{equation}
\end{widetext}
where $\zeta=(\frac{\theta_p}{B\theta_0})^C$, and
\begin{equation}\label{Delta-esferica}
    \Delta=\delta\beta_s^2-\left(\delta\beta_{20}\cos\phi-\delta\beta_{10}\sin\phi+\frac{\eta |\vec{\theta}_E|\sin2\phi}{\sqrt{1-\eta\cos2\phi}}\right)^2.
\end{equation}

As we have pointed out, image formation is characterized by inequality \eqref{image-condition}. From this condition we can analytically analyze, for some particular cases, the regions in the lens plane where we will have images in terms of source and lens parameters. The first thing that we can notice is that a spherical plasma profile will not have an effect on the angular position of the images since $\Delta$ does not depend on the observation frequency or on the electronic distribution on the plasma. So the effect of the plasma in the images position will be completely in the radial direction.

Let us first consider the case without ellipticity ($\eta = 0$) for three different arrangements of the source position. For the case in which the source is horizontally aligned in the plane of the source, that is, for sources located along the line characterized by $\delta\beta_{10}\neq0$ and $\delta\beta_{20}=0$, we see from \eqref{Delta-esferica} that the images in the lens plane will be restricted to regions $\sin^2\phi\leq (\delta\beta_s/\delta\beta_{10})^2$, with $0\leq\phi\leq 2\pi$. On the other hand, for vertically aligned sources, that is, for sources located along the line $\delta\beta_{10}=0$ and $\delta\beta_{20}\neq 0$, the images will be restricted to regions where $\cos^2\phi\leq(\delta\beta_s/\delta\beta_{20})^2$, with $0\leq\phi\leq 2\pi$. Finally, within the non-ellipticity case we can also obtain an explicit condition for the images position for sources located along the diagonal and anti-diagonal in the source plane characterized by $\delta\beta_{10}=\pm\delta\beta_{20}\neq 0$. In such case images position in the lens plane will be characterized by the condition $\sin^2(\pi/4\mp\phi)\leq\frac{1}{2}(\delta\beta_s/\delta\beta_{20})^2$, with $0\leq\phi\leq 2\pi$.

On the other hand, we can also analyze the case where the ellipticity of the lens is extremely small ($\eta\ll 1$)  but where the source is located at the origin of the source plane, that is, $\delta\beta_{10}=\delta\beta_{20}=0$. In this case images formation will occur in the region of the lens plane characterized by $\sin^2(2\phi)\leq (\frac{\delta\beta_s}{\eta\theta_E})^2$, with $0\leq\phi\leq 2\pi$.

In Fig. \ref{hhh} we graphically compare the perturbative with the numerical method with this particular plasma model for the following lens configuration. We consider a gravitational lens described by the parameters: $n_0=60$cm$^{-3}$, $r_p=1$kpc, $\sigma_c=180$km/s, $\eta=0$.$3$ at an observation frequency $\nu=80$Mhz. The angular radius of the Einstein ring for this configuration is $\theta_E=0$.$555$arcsec (green line) while $\theta_E^p=0$.$517$arcsec (black line). The source parameters are (radius and position): $\delta\beta_s=0$.$06\theta_E$, $\delta\beta_{10}=0$.$08\theta_E$, $\delta\beta_{20}=0$.$0$. As mentioned in the introduction, we assume that the lens and the source are located at $z_l=0$.$04$ and $z_s=0$.$1$, respectively.  
In this figure we show the solution of the lens equation for pure gravity case in red lines, while in blue lines we show the solution for the case with plasma. In this particular case we see that the plasma does not change the multiplicity of images or their morphology and, in particular, we can see that perturbative solution fully coincides with the numerical one (gray line) for a single iteration. However, we see that for this configuration, the main effect of the plasma is the shift  of images position in the radial direction towards the center of the lens.  It is worth mentioning that in radio frequency observations, although they are focused on certain centered observation frequencies, they typically have a bandwidth, which can result in blurred images. However, for example, the LOFAR observatory has a bandwidth of approximately 3 MHz for observations at 140 MHz or less \cite{refId0}, so the position and shape of the images will only change slightly. In this work, which aims to study the main effects of plasma, we will omit this kind of consideration. Nevertheless, one could use the same analytical formulas developed here to describe the change in position and blurring of the images.

Finally, it is important to point out that all numerical solutions of the lens equation as well as the numerical computation of critical and caustic curves obtained in Sec. \ref{galaxy_modelling}, \ref{sec4} and \ref{caustic-critical} were carried out with the multi-purpose open-source gravitational lensing Lenstronomy package\cite{BIRRER2018189}, which was suitable modified to include the plasma models used in this work.

\subsection{Exponential model for an edge-on plasma disk}

Although a spherically symmetric electron distribution in the plasma allows us to show the basic effects of the plasma's influence on image formation, it may not be a sufficiently realistic model, so it is also worth studying electronic distributions that respect other symmetries. In this subsection we will consider a electron distribution with azimuthal symmetry about the $z^\prime$-coordinate axis. In this case it is convenient to work with cylindrical coordinates in such a way that the electron density in the plasma only depends explicitly on the $z^\prime$ and $r_c$ coordinates. Recall that the cylindrical radial coordinate $r_c$ is defined on the $x^\prime y^\prime$ plane perpendicular to the lens plane as shown in Fig. \ref{coord}. We choose an exponential decay in the radial direction and we will make two different choices for the behavior of the electron density in the $z^\prime$ direction. Thus we will consider an electron density of the form,
\begin{equation}
    n_e(r_c,z^\prime)=n_0 e^{-r_c/r_p}f(z^\prime),
\end{equation}
with $r_p$ a free parameter.
This kind of models, more realistic than those spherically symmetric (see equation \eqref{exponential_density}) have also been considered in the past, obtaining several estimates for the average electron density $n_0$ and the characteristic radius $r_p$ from the study of the ion distribution in the H II regions for various galaxies. In \cite{lima} values of $n_0 = 500\text{cm}^{-3}$ and $r_p = 8$kpc were obtained for the galaxy NGC 1232, while in \cite{Davies_2021} values for $n_0$ were estimated in the range of $\approx [30 - 260]\text{cm}^{-3}$, based on the study of more than 600 galaxies from the KMOS and SAMI surveys.

In this case the electron column density reads,
\begin{equation}
\begin{aligned}
    N_e(\theta,\varphi)&=\int_{-D_l}^{D_{ls}}n_e(r_c,z')dx' \\
    &=n_0 f(z') \int_{-D_l}^{D_{ls}} e^{-\frac{\sqrt{x'^2+(D_l\theta\cos\varphi)^2}}{r_p}}dx' \\
    &=n_0 r_p f(z') \int_{-D_l/r_p}^{D_{ls}/r_p} e^{-\sqrt{\tilde{x}^2+(\frac{D_l}{r_p}\theta\cos\varphi)^2}}d\tilde{x},
\end{aligned}
\end{equation}
where $\tilde{x}=x'/r_p$. As this integral cannot be performed analytically we choose to approximate it by a exponential function as follows,
\begin{equation}
    N_e(\theta,\varphi)=n_0 r_p f(z') \left( A e^{-(\theta |\cos\varphi| /B\theta_0)^C} \right),
\end{equation}
where $\theta_0=r_p/D_l$, and $A, B$ and $C$ are dimensionless parameters. As we have mentioned in the previous subsection these parameters depend on the particular choice of $r_p$. Therefore the plasma potential for this profile is given by,
\begin{equation}
    \psi_{\text{plasma}}(\vec{\theta},\omega)=\psi_\omega^2 e^{-\left(\frac{\theta|\cos\varphi|}{B\theta_0}\right)^C} f(z'),
\end{equation}
where 
\begin{equation}
    \psi_\omega^2=\frac{D_{ls}}{D_s D_l}\frac{2\pi c^2}{\omega^2(1+z_l)^2} A r_e n_0 r_p.
\end{equation}

Unlike of the case considered in the previous subsection, the resulting plasma potential is not circularly symmetric. Therefore, unlike the spherically symmetric model we described above, in this case the plasma potential acts as a perturbation in the lens equation. In such a way that the solution of the zero-order lens equation $\theta_p$ given by \ref{lens-zero} will coincide with the radius of Einstein's ring, that is, $\theta_p = \theta_E$. Therefore, we have the following expressions for the umperturbed and perturbed potentials (considering again the SIE gravitational potential as a model of the dark matter halo):
\begin{widetext}
 \begin{eqnarray}
 \psi_{\text{tot}}(\vec{\theta})&=&\psi_{\text{grav}}(\vec{\theta})+\psi_{\text{plasma}}(\vec{\theta},\omega),\\
    \psi_0(\vec{\theta})&=& \psi_{\text{grav}}(\vec{\theta})|_{\eta=0}=\theta_E \;\theta,\\
    \delta{\psi}(\vec{\theta})&=&\psi_{\text{tot}}(\vec{\theta})-\psi_0(\theta)= \theta_E \;\theta(\sqrt{1-\eta\cos 2\phi}-1)+\psi_{\text{plasma}}(\vec{\theta},\omega).   
\end{eqnarray}
\end{widetext}

Finally, we will choose two different kinds of decay along the $z'$-direction, for both positive and negative values of this coordinate. First an exponential decay and then a Gaussian one. In both cases the idea is to simulate that the electrons are mostly distributed along the $x'y'$ plane, that is, the decay in the $z$ direction must be faster than along the $xy$ plane. As we did in the previous case, in order to describe the images in the lens plane, we will express the lens potential in terms of the coordinates $b = D_l \theta$ and $\varphi$ fitted to the plane of the lens.

\subsubsection{Exponential decay in $z'$-direction}

 For this model, we consider the following $f(z)$ function,
\begin{equation}
    f(z')=e^{-|z'|/z_0}=e^{-\theta |\sin\varphi|/\theta_z},
\end{equation}
where $z_0$ is a parameter, $\theta_z=z_0/D_l$, while the plasma potential will be given by,
\begin{equation}
    \psi_{\text{plasma}}(\vec{\theta},\omega)=\psi_\omega^2 e^{-\left(\frac{\theta|\cos\varphi|}{B\theta_0}\right)^C} e^{-\frac{\theta|\sin\varphi|}{\theta_z}}.
\end{equation}
Thus, the solution of the lens equation is given by the following expression,
\begin{equation}
\begin{aligned}
    \delta\theta^\pm =& \delta\beta_{10}\cos\varphi+\delta\beta_{20}\sin\varphi+\theta_E(\sqrt{1-\eta\cos 2\varphi}-1)\\
    &-\frac{\psi_\omega^2}{\theta_p}e^{-\chi}\bigg(\chi+(C-1)(\frac{\theta_p}{B\theta_0})^C |\cos\varphi|^C\bigg)\pm\sqrt{\Delta},
\end{aligned}    
\end{equation}
where $\chi=(\frac{\theta_p}{B\theta_0})^C|\cos\varphi|^C+\frac{\theta_p}{\theta_z}|\sin\varphi|$, and
\begin{widetext}
\begin{equation}
\begin{aligned}
    \Delta=&\delta\beta_s^2-\bigg( \delta\beta_{20}\cos\varphi-\delta\beta_{10}\sin\varphi+\frac{\eta\theta_E\sin 2\varphi}{\sqrt{1-\eta\cos 2\varphi}}+\frac{\psi_\omega^2}{\theta_p}e^{-\chi} \left(C(\frac{\theta_p}{B\theta_0})^C |\cos\varphi|^C \tan\varphi  
    -\frac{\theta_p}{\theta_z}|\sin\varphi|\cot\varphi \right)\bigg)^2.
\end{aligned}    
\end{equation}
\end{widetext}

In this way we obtain an analytical solution to determine the position of the images in the lens plane. Although in this case an analytical study can also be carried out on the images position for different locations of the source as we did for the spherical model, the expressions that result from such analysis do not provide as much clarity as in the previous case and for this reason we decided directly face a graphic analysis of the images. What we can say from the analytical solution is that, unlike the spherically symmetric case, the effect of the plasma in the images will occur not only in the radial direction but also in its angular position, since in this case the function $\Delta$ depends on the plasma parameters.

\begin{figure*}
\centering
\includegraphics[width=\textwidth]{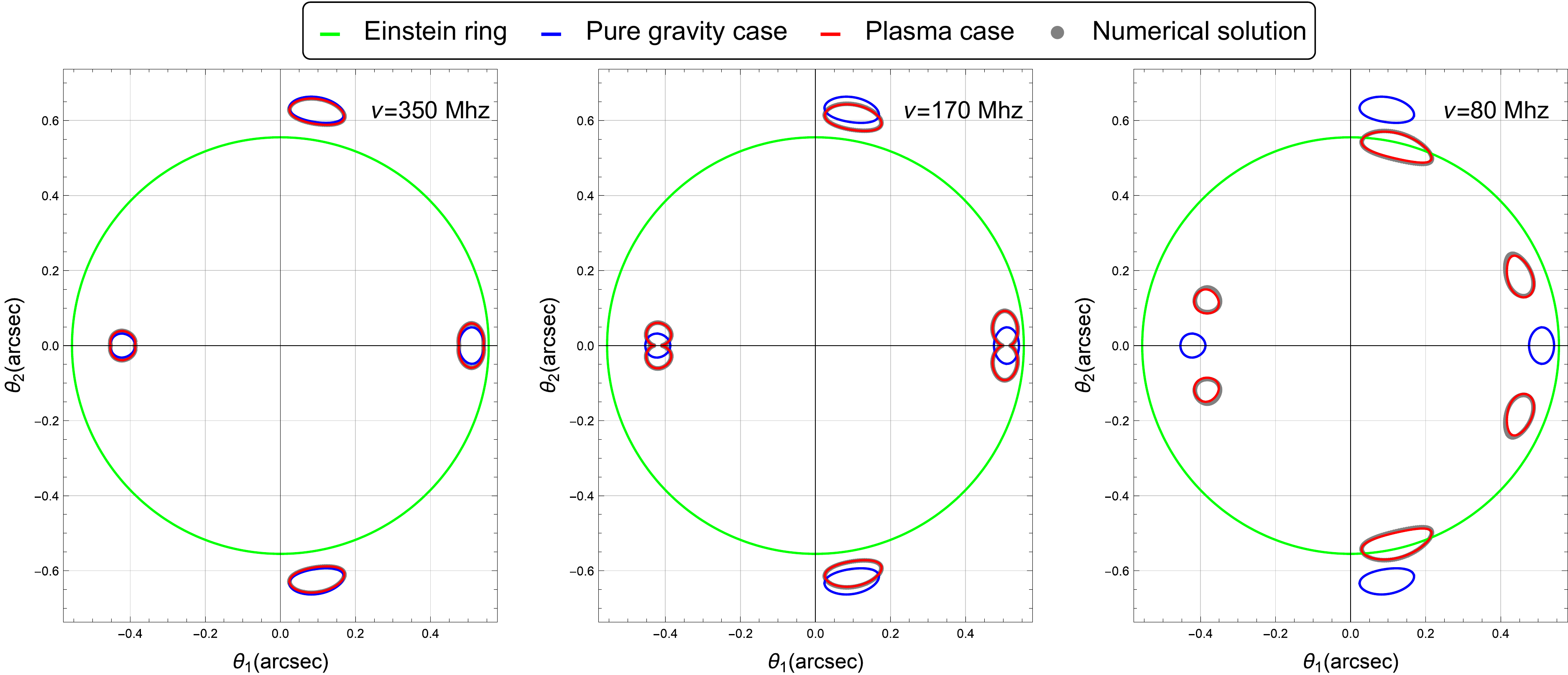}
\caption{SIE model for a plasma disk with exponential decay in the $z$ direction with parameters $n_0=10$cm$^{-3}$, $r_p=10$kpc, $z_0=1$kpc, $\sigma_c=180$km/s , $\eta=0$.$3$, $\theta_E=0.555$arcsec (green line) for a circular source with radius $\delta\beta_s=0.06\theta_E$ centered at $\delta\beta_{10}=0.08\theta_E$, $\delta\beta_{20}=0.0$.}
\label{hhh1}
\end{figure*}

In  Figure \ref{hhh1} we consider as an example a circular source with radius $\delta\beta_s=0.06\theta_E$ centered at $\delta\beta_{10}=0.08\theta_E$, $\delta\beta_{20}=0.0$, and parameters $n_0=10$cm$^{-3}$, $r_p=10$kpc, $z_0=1$kpc, $\sigma_c=180$km/s , $\eta=0.3$, $\theta_E=0$.$555$arcsec (green line). The fitting parameters are given by $A=2.00004\pm 0.00002$, $B=1.16\pm 0.01$, $C=1.719\pm 0.006$. In blue lines we see the solution of the lens equation for the case of pure gravity while in red we plot the perturbative solution with plasma and compare it with the numerical integration of the lens equation (gray line) in order to corroborate the accuracy of the perturbative approach.

In this case we plot the images for three different observation frequencies, from left to right: 350Mhz, 170Mhz and 80Mhz. For the higher frequency that we are considering the plasma effect is hardly distinguishable from pure gravity case which is the expected behaviour. For a frequency lower than 170Mhz, we see that the images that are further away from the horizontal axis, which coincides with the $z=0$ axis in the lens plane, tend to get closer to the center of the lens, a change is also seen in the angular position of the images with respect to the case of pure gravity. On the other hand, we see that the images close to the horizontal axis, which in turn coincides with the plasma disk, tend to separate, forming for low frequencies four images instead of the two that appeared in the gravity pure case. This is because of due to the divergent property of the plasma lensing, some rays of light deviate above and others below the horizontal axis. These facts are evidenced in the last plot for a frequency of 80Mhz, showing in this case that plasma can not only interfere with the morphology of the images but also with their multiplicity.

Finally, we point out that these plots were obtained with a single iteration of our method and in comparison with the numerical solution we see that the method is quite accurate at least for this studied configuration, although as it can be seen, it is less accurate for lower and lower frequencies.

\subsubsection{Gaussian-like decay in $z'$-direction}

Let us consider the following $f(z')$ function,
\begin{equation}
    f(z')=e^{-(z'/z_0)^2}=e^{-(\theta\sin\varphi/\theta_z)^2},
\end{equation}
where as before, $\theta_z=z_0/D_l$, while the plasma potential will be given by,
\begin{equation}
    \psi_{\text{plasma}}(\vec{\theta},\omega)=\psi_\omega^2 e^{-\left(\frac{\theta|\cos\varphi|}{B\theta_0}\right)^C} e^{-\left(\frac{\theta\sin\varphi}{\theta_z}\right)^2}.
\end{equation}
Thus, the solution of the lens equation is given by the following expression,
\begin{widetext}
\begin{equation}
\begin{aligned}
    \delta\theta^\pm =& \delta\beta_{10}\cos\varphi+\delta\beta_{20}\sin\varphi+\theta_E(\sqrt{1-\eta\cos 2\varphi}-1)-\frac{\psi_\omega^2}{\theta_p}e^{-\xi}\bigg(\xi+(C-1)(\frac{\theta_p}{B\theta_0})^C |\cos\varphi|^C+\frac{\theta_p^2}{\theta_z^2}\sin^2\varphi\bigg)\pm\sqrt{\Delta},
\end{aligned}    
\end{equation}
where $\xi=(\frac{\theta_p}{B\theta_0})^C|\cos\varphi|^C+\frac{\theta_p^2}{\theta_z^2}\sin^2\varphi$, and
\begin{equation}
    \Delta=\delta\beta_s^2-\Bigg( \delta\beta_{20}\cos\varphi-\delta\beta_{10}\sin\varphi+\frac{\eta\theta_E\sin 2\varphi}{\sqrt{1-\eta\cos 2\varphi}}
    +\frac{\psi_\omega^2}{\theta_p}e^{-\xi} \left(C(\frac{\theta_p}{B\theta_0})^C |\cos\varphi|^C \tan\varphi  -2\frac{\theta_p^2}{\theta_z^2}\cos\varphi\sin\varphi \right)\Bigg)^2.   
\end{equation}
\end{widetext}
We see again that the plasma will have effects on both the radial and angular position of the images in the lens plane since $\Delta$ also depends on the parameters of the plasma.

\begin{figure*}
\centering
\includegraphics[width=\textwidth]{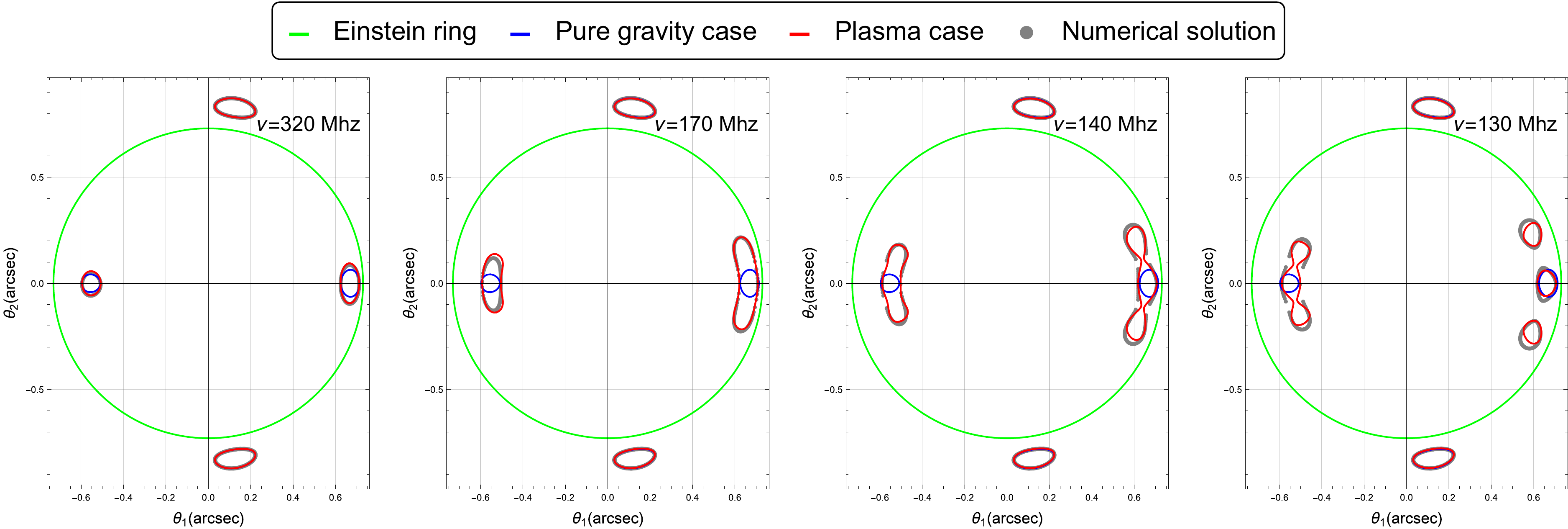}
\caption{
SIE model for a plasma disk with Gaussian decay in the z direction with parameters $n_0=40$cm$^{-3}$, $r_p=10$kpc, $z_0=1$kpc, $\sigma_c=180$km/s , $\eta=0$.$3$, $\theta_E=0$.$730$arcsec (green line) for a circular source with radius $\delta\beta_s=0.06\theta_E$ centered at $\delta\beta_{10}=0.08\theta_E$, $\delta\beta_{20}=0. 0$.}
\label{hhh2}
\end{figure*}

In Figure \ref{hhh2} we consider as an example a circular source with radius $\delta\beta_s=0.06\theta_E$ centered at $\delta\beta_{10}=0.08\theta_E$, $\delta\beta_{20}=0. 0$, and parameters $n_0=40$cm$^{-3}$, $r_p=10$kpc, $z_0=1$kpc, $\sigma_c=180$km/s , $\eta=0.3$, $\theta_E=0.730$arcsec (green line). The fitting parameters in this case are $A=2.008\pm 0.002$, $B=1.590\pm 0.005$, $C=1.430\pm 0. 008$. In blue lines we see the solution of the lens equation for pure gravity case while in red we plot the perturbative solution with plasma and compare it with the numerical integration of the lens equation (gray line) in order to corroborate the accuracy of the method.

In this case we plot the images for four different observation frequencies, from left to right: 320Mhz, 170Mhz, 140Mhz, 130Mhz. Again for high frequencies the effect of the plasma is quite weak. Although as we go to lower and lower observation frequencies we see a similar situation with the previous model. The images that are close to the horizontal axis $z'=0$ begin to separate but this time in three images each of them, while the two images that are further from the axis $z'=0$ remain practically unchanged with respect to the pure gravity case because, for this particular model, the decay along the $z'$ direction is much faster than in the previous model and therefore the influence of the plasma in these images is very slight. It is also for this reason that the perturbative solution resolves these images quite well and not so well those close to the $z'=0$ axis. In particular, we see that for the observation frequency of 130Mhz, the perturbative solution would need at least another iteration to be able to reproduce the images properly.

The effect produced by the plasma in the multiplicity of images is closely related to the effect produced by the plasma on the structure of caustic curves, and therefore also on the structure of critical curves. In section \ref{caustic-critical} we will study the effect of plasma on these kinds of curves, thus showing what causes this doubling in the images produced by plasma.

\subsection{Gaussian model: front view}

Let us now consider a plasma disk seen head-on with Gaussian decay both along and across in the plane perpendicular to the line of sight. In this case the electron density will be given as follows,
\begin{equation}
n_e(x',y',z')=n_0 e^{-\frac{y'^2+z'^2}{r_p^2}} e^{-\frac{x'^2}{z_0^2}}.
\end{equation}
Then, the projected electron density along the line of sight reads,
\begin{equation}
\begin{aligned}
    N_e(y',z')&=\int_{-D_l}^{D_{ls}}n_e(x',y',z') dx' \\
    &= n_0 e^{-\frac{y'^2+z'^2}{r_p^2}} \int_{-D_l}^{D_{ls}}e^{-\frac{x'^2}{z_0^2}} dx' \\
    &= n_0 z_0 e^{-\frac{y'^2+z^2}{r_p^2}} \int_{-D_l/z_0}^{D_{ls}/z_0}e^{-\tilde{x}^2} d\tilde{x},
\end{aligned}
\end{equation}
where $\tilde{x}=x/z_0$. In general, for the situations that we will be considering, the distances $D_l$ and $D_{ls}$ are of the order of Mpc (megaparsec) while $z_0$, the characteristic scale of the plasma disk, is of the order of kpc (kiloparsec). Added to the fact that the integrand $e^{-\tilde{x}^2}$ decays fast enough, we can replace the limits of integration $D_l/z_0$ and $-D_{ls}/z_0$ in the above integral by the asymptotic values $\infty$ and $-\infty$, respectively; and thus obtain a good approximation of it. In addition, this will allow us to solve the integral analytically. Therefore,
\begin{equation}
\begin{aligned}
    N_e(y',z') &\approx n_0 z_0 e^{-\frac{y'^2+z'^2}{r_p^2}} \int_{-\infty}^{\infty}e^{-\tilde{x}^2} d\tilde{x} \\
    &= n_0 z_0 e^{-\frac{y'^2+z'^2}{r_p^2}} \sqrt{\pi}.
    \end{aligned}
\end{equation}
Rewriting $N_e$ in terms of the angular coordinate $\theta$ using the relations \eqref{coord-strong} and \eqref{coord-strong-1} we obtain,
\begin{equation}
    N_e(\theta) = n_0 z_0 e^{-\theta/\theta_0} \sqrt{\pi},
\end{equation}
where $\theta_0=r_p/D_l$. Finally the plasma potential will be given as follows,
\begin{equation}\label{expo_plasma-0}
    \psi_{\text{plasma}}(\vec{\theta},\omega)=\psi_\omega^2 e^{-\theta/\theta_0},
\end{equation}
where
\begin{equation}\label{expo_plasma-1}
    \psi_\omega^2=\frac{D_{ls}}{D_s D_l}\frac{2\pi c^2}{\omega^2(1+z_l)^2} r_e n_0 z_0\sqrt{\pi}.
\end{equation}
Note that this plasma profile will have a similar effect to the spherically symmetric plasma profile we consider in \eqref{exponential_density} because in both cases the projected electron density $N_e$ is axially symmetric with respect to the line of sight. Indeed, we can see the similarity of the plasma potentials if we compare the equations \eqref{expo_plasma-0} and \eqref{expo_plasma-1} with the equations \eqref{spherical-0} and \eqref{spherical-1}. For this reason, in the remainder of the article we will not mention this profile because any analysis that we could carry out is somehow contained in the analysis carried out for the spherically symmetric density profile given by \eqref{exponential_density} .

Lastly, the effect that a plasma disk like the one we are considering, for an arbitrary orientation with respect to the line of sight, has on image formation was recently discussed and can be consulted in \cite{Tomas-tesina,Tomas-work}.

\section{Comparison for several iterations}\label{sec4}

Because this is a perturbative method, it is expected that for some situations the solutions obtained through it are too far from the exact solutions. In this Section we will call exact solutions to those obtained numerically since we can obtain them with a high degree of precision, even though they are not strictly so. This situation where the perturbative method is not precise enough can be reached in various circumstances, either because the ellipticity of the lens is very high or because the source is centered too far from the line of sight or its radius is too large. Or a combination of them. These situations have been studied both in Alard's original work and in subsequent works \cite{Alard:2007ya,Dumet-Montoya:2013yqp,Habara11}.

On the other hand, the plasma will also influence the accuracy of the perturbative method, since for plasma potentials without spherical symmetry it will intervene as a perturbation of the lens potential in pure gravity, such is the case, for example, of the plasma disk seen from side that we analyzed in the previous section. But also the spherically symmetric plasma potentials will have an effect on the accuracy of the method since, as we can see, the first order solution given by \eqref{ec-lente-primer-orden} must be evaluated in $\theta_p$ which corresponds with the zero order solution of the lens equation that is affected by the plasma. Obviously, the influence of the plasma on the accuracy of the perturbative method will be greater as the electron density increases as well as for lower and lower observation frequencies.

As we have seen, in  Sec. \ref{N-iteraciones} we introduced an iterative correction of the perturbative method in order to address those situations where the perturbative method (with a single iteration) is not accurate enough to reproduce the exact solutions.

\begin{figure*}
\centering
\includegraphics[width=\textwidth]{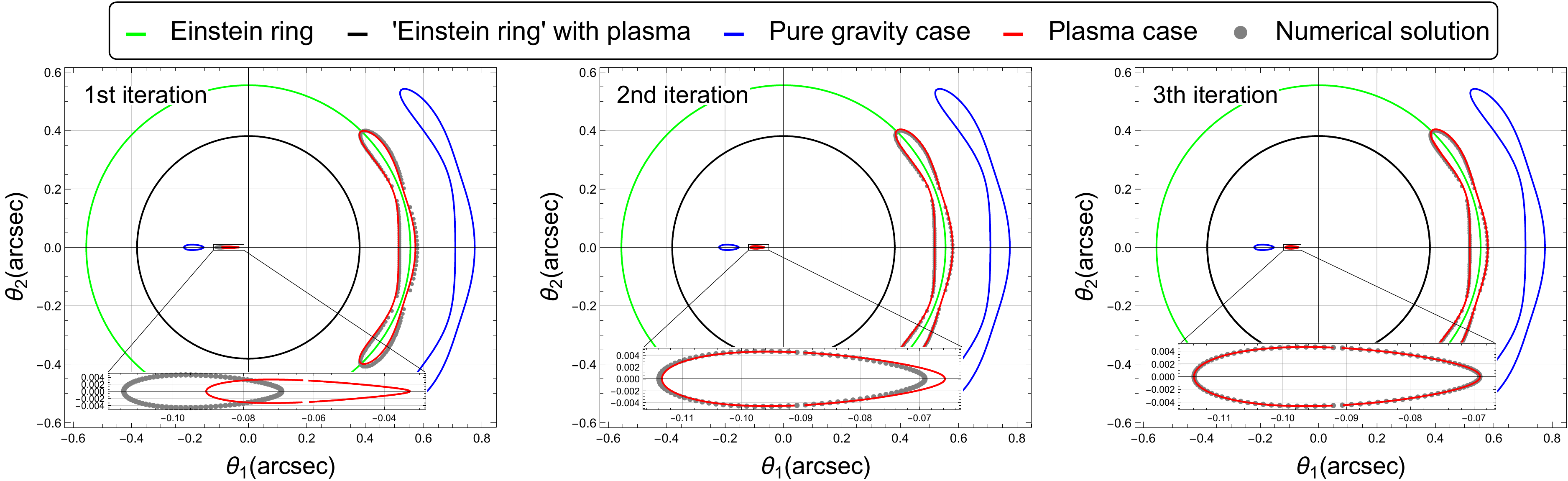}
\caption{SIE model with spherical plasma and exponential decay with pameters $\nu=80$Mhz, $n_0=300$cm$^{-3}$, $r_p=1$kpc, $\sigma_c=180$km/s, $\eta=0.3$, $ \theta_E=0.555$arcsec (green line), $\theta_E^p=0.381$arcsec (black line) for a circular source with radius $\delta\beta_s=0.06\theta_E$ centered at $\delta\beta_{10}=0.5\theta_E$, $\delta\beta_ {20}=0.0$. Comparison for the first three iterations.}
\label{hhh3}
\end{figure*}

In Fig. \ref{hhh3} we return to the spherical plasma model for the following configuration: $\nu=80$Mhz, $n_0=300$cm$^{-3}$, $r_p=1$kpc, $\sigma_c=180$km/s, $\eta=0.3$, $ \theta_E=0.555$arcsec (green line), $\theta_E^p=0.381$arcsec (black line). Source parameters (radius and position where it is centered): $\delta\beta_s=0.06\theta_E$, $\delta\beta_{10}=0.5\theta_E$, $\delta\beta_ {20}=0.0$. The fitting parameters are: $A=2.003\pm 0.002$, $B=1.55\pm 0.01$, $C=1.47\pm 0.01$. In this figure we see the images formation for the first three iterations of the perturbative method. In gray is the exact solution (or numerical solution itself). Both for the case of pure gravity (blue lines) and for the case with plasma (red lines) we see that two images are formed: on the right a gravitational arc that is a tangential deformation of the solution to order zero, and on the left side a radially deformed image, which we enlarged for better visualization. The corrections introduced by the second and third iterations are clearly evident in the enlarged region where we see how the accuracy of the perturbative method improves substantially, while the correction in the gravitational arc does not seem to change much to the naked eye. In this case the effect of the plasma is relevant in the accuracy of the method both because we are working with a relatively low observation frequency of 80Mhz and also because we are considering a relatively high electron density of $300\text{cm}^{-1}$ in comparison with the cases that we have been analyzing up to now.

\begin{figure}
\centering
\includegraphics[width=\columnwidth]{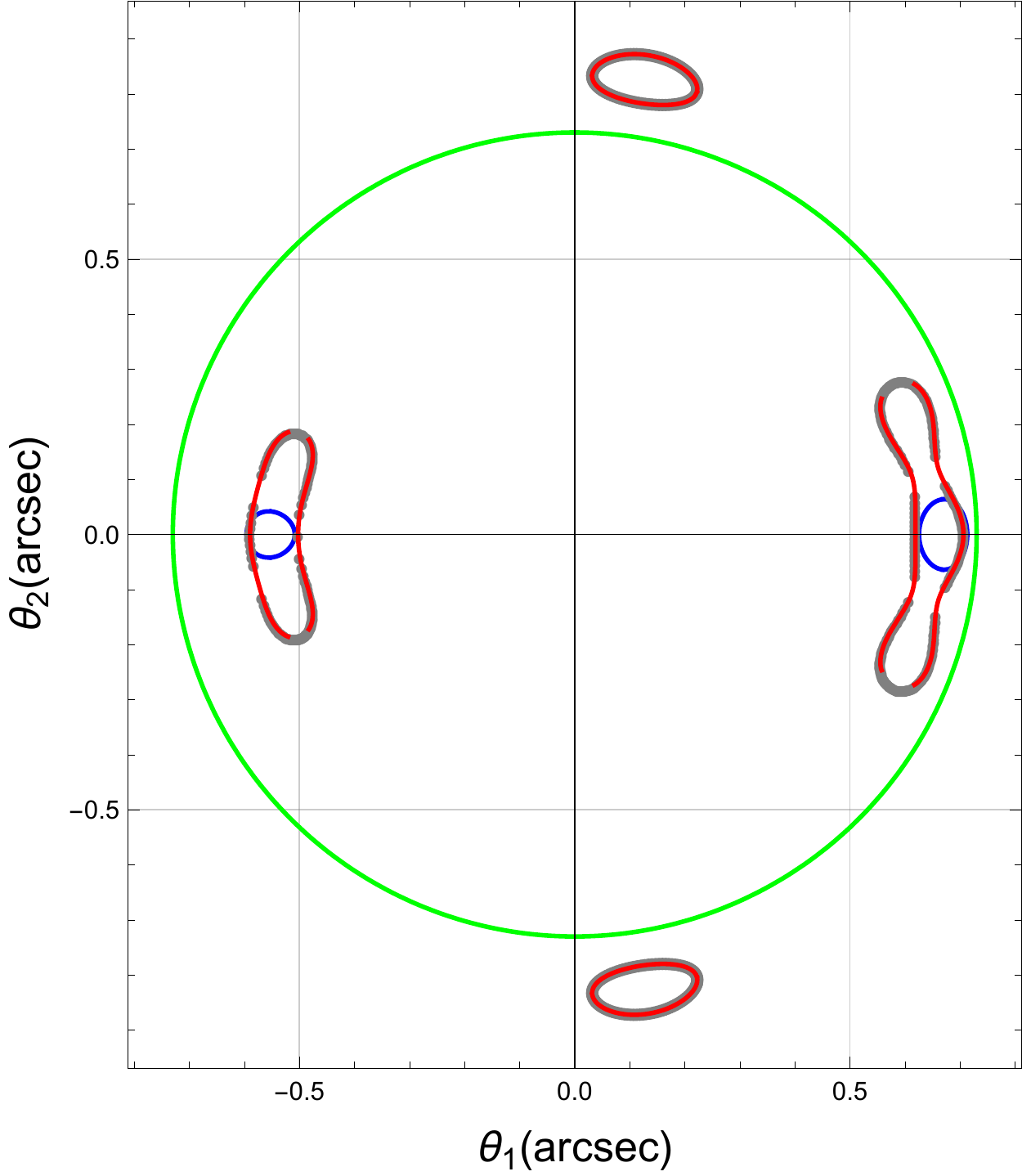}
\caption{SIE model for a plasma disk with Gaussian decay in $z$ direction, repeating the configuration of Fig. \ref{hhh2} with observation frequency $\nu = 140\text{Mhz}$. Third iteration.}
\label{hhh4}
\end{figure}

\begin{figure}
\centering
\includegraphics[width=\columnwidth]{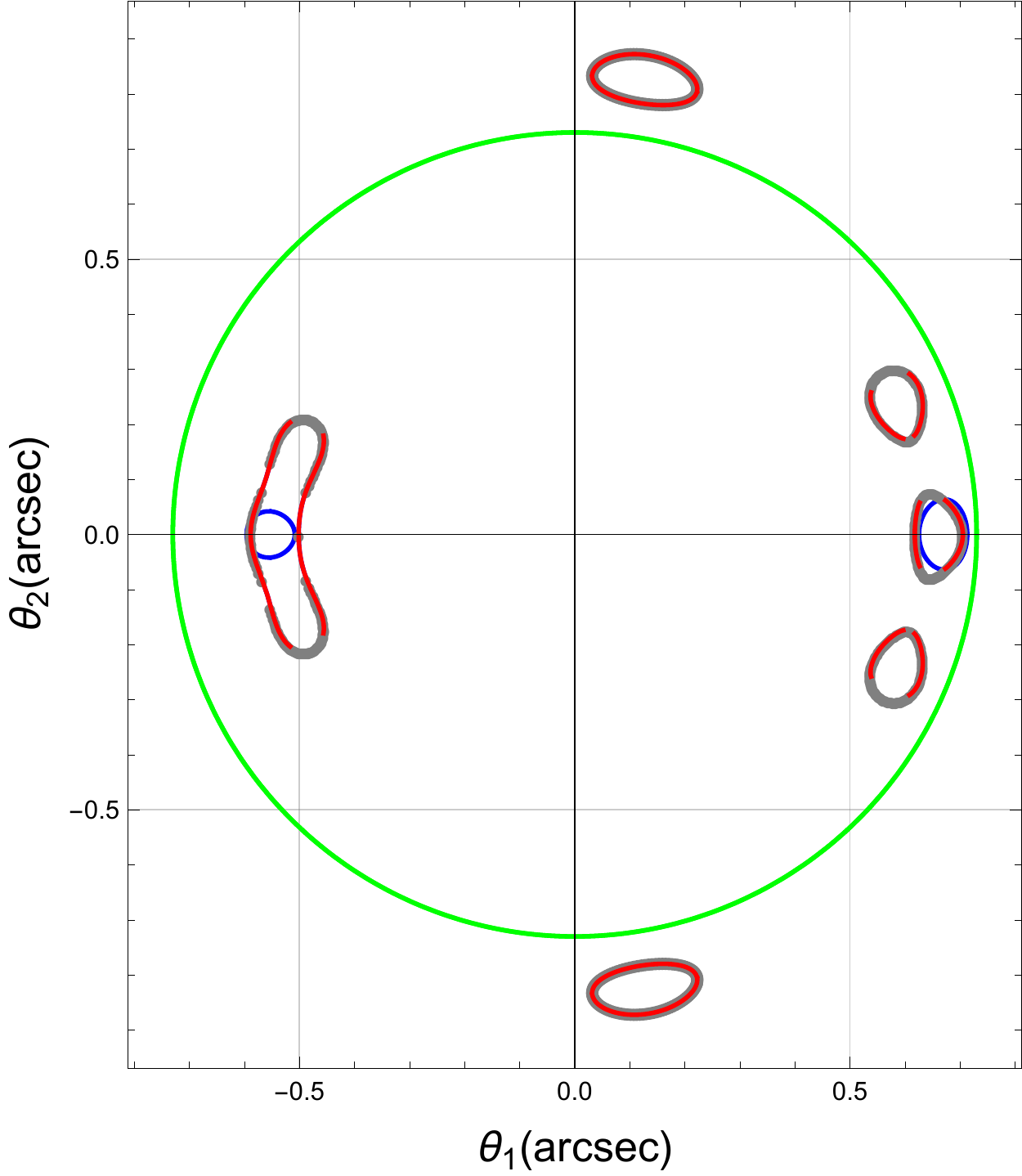}
\caption{SIE model for a plasma disk with Gaussian decay in $z$ direction, repeating the configuration of Fig. \ref{hhh2} with observation frequency $\nu = 130\text{Mhz}$. Third iteration.}
\label{hhh5}
\end{figure}

On the other hand, we also analyze the iterative corrections of the perturbative method for the plasma disk seen from the side with Gaussian decay in the $z$ direction. In this case we repeat the same configuration of parameters that we used in Fig. \ref{hhh2} and in particular we are going to concentrate on the last two subfigures on the right that correspond to observation frequencies of 140Mhz and 130Mhz, respectively. In both cases we will see how a third iteration of the perturbative method produces significant corrections of the perturbative method. These corrections can be seen in Fig. \ref{hhh4} for the 140Mhz frequency, and Fig. \ref{hhh5} for the 130Mhz frequency. Higher order iterations do not show significant corrections.

In this way we graphically show for some particular examples how the iterative corrections of the perturbative method are useful to reproduce the position of the images more faithfully. We highlight that in all cases corrections of order four or higher were not necessary. On the other hand, although it is possible to show how the analytical solutions are for the different iterations, these expressions are quite cumbersome and lack any illustrative character, and for this reason we decided to carry out only a graphical analysis of them.

\section{Critical and caustic curves}\label{caustic-critical}

As we saw in Sec. \ref{galaxy_modelling}, for certain plasma profiles, in particular for a plasma disk seen from the side with both exponential and Gaussian decay in perpendicular direction to the disk, the plasma has a clear effect on the multiplicity of images (see Fig. \ref{hhh1} and Fig. \ref{hhh2}). Far from being the goal of this section to establish a general criterion or necessary conditions that allow us to predict the number of images given a certain plasma profile, we are going to analyze the effect of plasma on the critical and caustic curves associated with Fig. \ref{hhh2}, which corresponds to the model with Gaussian decay in the $z$ direction. The plasma will have a similar effect on the critical and caustic curves associated with Fig. \ref{hhh1} although in this case, since the lens potential is not differentiable at $z=0$, they must be carefully calculated near this area.

The critical curves, which are those curves defined from the condition $J=0$ in the lens plane, where $J=\text{det}\frac{\partial\vec{\beta} }{\partial\vec{\theta}}$ is the determinant of the Jacobian matrix, are of significant importance because they are related to some of the most notorious effects of gravitational lensing theory: image magnification and multiplicity of them (the latter, characteristic of strong gravitational lenses). Due to the magnification effect, the image of an infinitesimally small source located at position $\vec{\theta}$ will magnify by a factor $|\mu(\vec{\theta})|$ where $\mu$ is known as magnification (or point magnification to be precise) and is defined by $\mu=\frac{1}{J}$. In this way, we see that the critical curves are those regions in the lens plane where the magnification of the images is infinite. This divergence indicates that the geometric optics approach fails in this region. However, when dealing in practice with extended sources, the magnification is calculated by averaging the point magnification on the source and, in turn, weighing it and normalizing it by its surface brightness. For a detailed discussion of this topic we refer to \cite{Schneider92}.

Let's assume, for example, that we have a spherical gravitational potential $\psi_0$ and a small perturbation $\delta\psi$ associated with the plasma medium. In this situation, in the perturbative approach, using \eqref{eq:pertprinci} we have
\begin{widetext}
\begin{equation}
	\begin{split}
		J = \frac{1}{\theta} \Bigg[
		\Big[  1 - \frac{\partial^2 (\psi_0 + \delta \psi)}{\partial \theta ^2}   \Big] 
		\Big[ \theta  - \frac{\partial(\psi_0 + \delta \psi)}{\partial \theta} - \frac{1}{\theta} \frac{\partial^2 (\psi_0 + \delta \psi)}{\partial \phi ^2}  \Big]  \\
		- \frac{1}{\theta} \Big[ \frac{1}{\theta} \frac{\partial (\psi_0 + \delta \psi)}{\partial\phi} - \frac{\partial^2 (\psi_0 + \delta \psi)}{\partial\theta\partial\phi}   \Big]^2
		\Bigg]\Biggr\rvert_{\theta = \theta_E + \delta \theta} \, .~
	\end{split}
\end{equation}
\end{widetext}
By doing a linear approximation in $\delta\theta$ and $\delta\psi$
and using \begin{gather}
	\frac{\partial\psi_0}{\partial\theta}\Bigr\rvert_{\theta = \theta_E} = \theta_{E} \, , \\
	\frac{\partial\psi_0}{\partial\phi} = 0 \, ,\\
	\frac{1}{\theta + \delta \theta} = \frac{1}{\theta} - \frac{\delta \theta}{\theta ^2} + \mathcal{O}(\delta \theta ^2) \, ,
\end{gather}
we finally obtain
\begin{gather}
	\label{jper}
	 J = \frac{1}{\theta} \left[ 1 - \frac{\partial^2 \psi _0}{\partial\theta^2}  \right] \left[  \delta\theta  \left( 1 - \frac{\partial^2 \psi _0}{\partial\theta^2}  \right) - \frac{\partial\delta \psi}{\partial \theta} - \frac{1}{\theta}\frac{\partial^2 \delta\psi}{\partial\phi^2}  \right]      
	\Biggr\rvert_{\theta = \theta_E} \, .
\end{gather}
From this relation it follows that if we consider 
that the plasma effect  introduced in 
$\delta\psi$ has circular symmetry,  the images can be 
demagnified with respect to the gravitational magnification 
for over-dense plasma regions (where 
$\frac{\partial\delta\psi}{\partial\theta}<0$). This case was 
analyzed in \cite{Er:2013efa}, finding that the ratio between 
the magnifications of different images does not change too 
much with respect to the case of considering only gravity, 
concluding therefore that the existence of plasma cannot 
account for the flux ratio anomaly.  The opposite effect is 
produced in under density regions. We refer to 
\cite{Er:2019jkg,Er:2017lue,Sun:2022ujt,Er:2022lad} for different situations that can be 
presented (see also \cite{Tomas-tesina} for a numerical study 
of the relative magnification between the different images). 
However, for non circular perturbative potentials, the effect 
of the plasma medium on the magnification of images could 
have a completely different behavior, depending strongly on 
the angular dependence in both $\theta$ and $\phi$ of the 
projected potential $\delta\psi$.  Note also that from the 
expression for the jacobian $J$, one infers that generically 
the total magnification is nonlinear. Therefore, in those 
cases, we cannot assign plasma and gravitational 
magnification values to each of the individual images.

In the following, instead of calculating the critical curves through the approximate formula eq.\eqref{jper} of $J$, we will show the graphs through numerical calculations without extra approximations. 

\begin{figure}
	\includegraphics[width=\columnwidth]{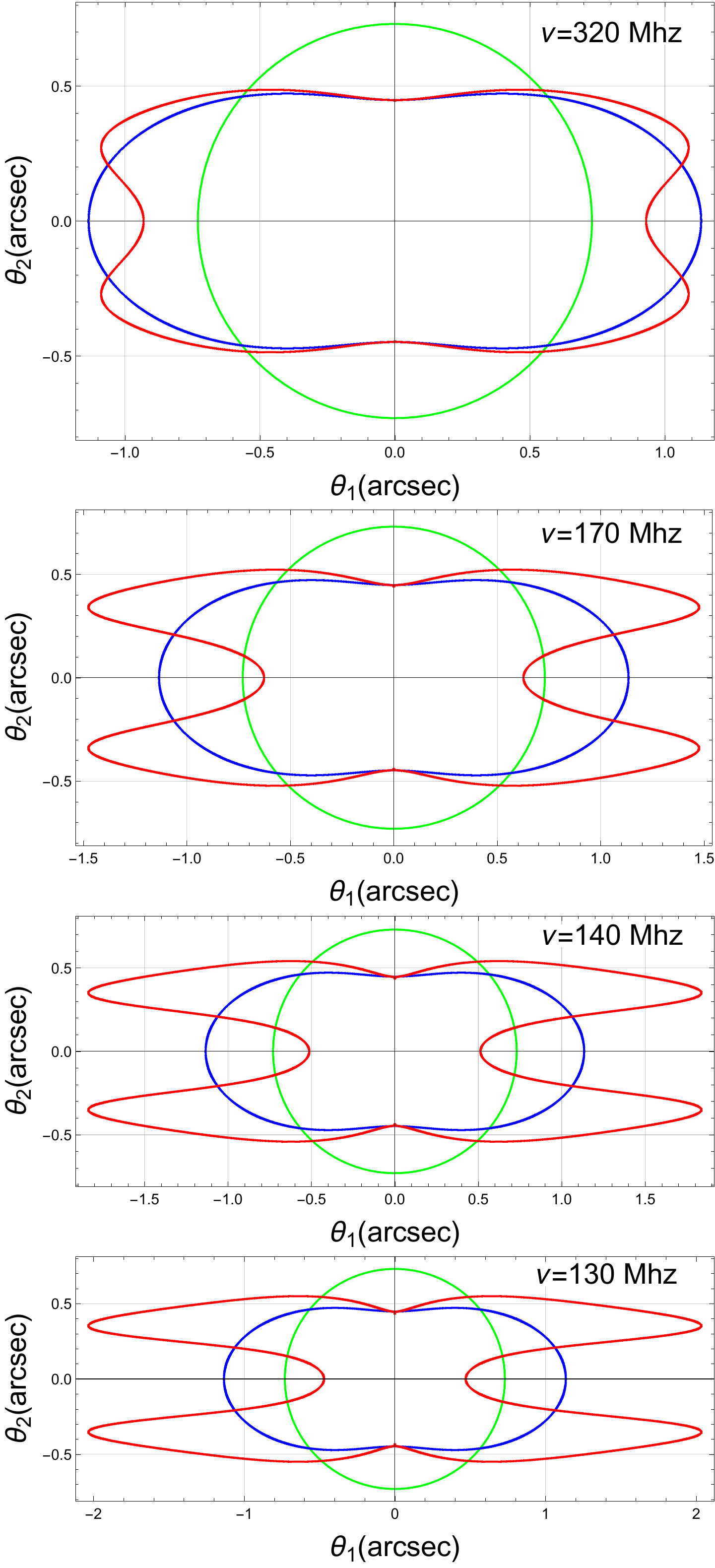}
    \caption{Critical curves associated to the SIE model of Fig. \ref{hhh2}, for a plasma disk with Gaussian decay in the z direction with parameters $n_0=40$cm$^{-3}$, $r_p=10$kpc, $z_0=1$kpc, $\sigma_c=180$km/s , $\eta=0$.$3$, $\theta_E=0$.$730$arcsec (green line) for a circular source with radius $\delta\beta_s=0.06\theta_E$ centered at $\delta\beta_{10}=0.08\theta_E$, $\delta\beta_{20}=0. 0$. Blue and red lines correspond to critical curves in the pure gravity case and in the plasma case, respectively.}
\label{critical}
\end{figure}

In Fig. \ref{critical} we see the critical curves associated with Fig. \ref{hhh2} for various observation frequencies. In green we plot the Einstein ring while in blue and red the critical curves for the pure gravity case and for the case with plasma, respectively. We can see that even for frequencies of 320MHz, the influence of the plasma is notorious and that its main influence, at least in this profile that we are considering, occurs along the horizontal axis defined by $z=0$ in the lens plane, because it coincides with the plasma disc seen from the side. Although such curves could also be obtained from a perturbative approach, we decided to numerically study both the critical and caustic curves in order to avoid any bias that a perturbative solution might introduce.

On the other hand, as we previously mentioned, the critical curves are also related to the multiplicity of images, although in an indirect way since the evaluation of the lens equation along these curves gives us what is known as caustic curves, and the relative position of the source with respect to these new curves is what will give us information about the multiplicity of the images. In other words, those sources that generate images located along critical curves in the lens plane are located along caustic curves in the source plane. A simple example of this is given in the case of an spherically symmetric lens and a point source, the latter aligned with the line of sight. The image produced in this case is an Einstein ring which in turn coincides with a critical curve, and therefore the caustic curve will be in this case the point where the source is located.

\begin{figure}[h!]
	\includegraphics[width=\columnwidth]{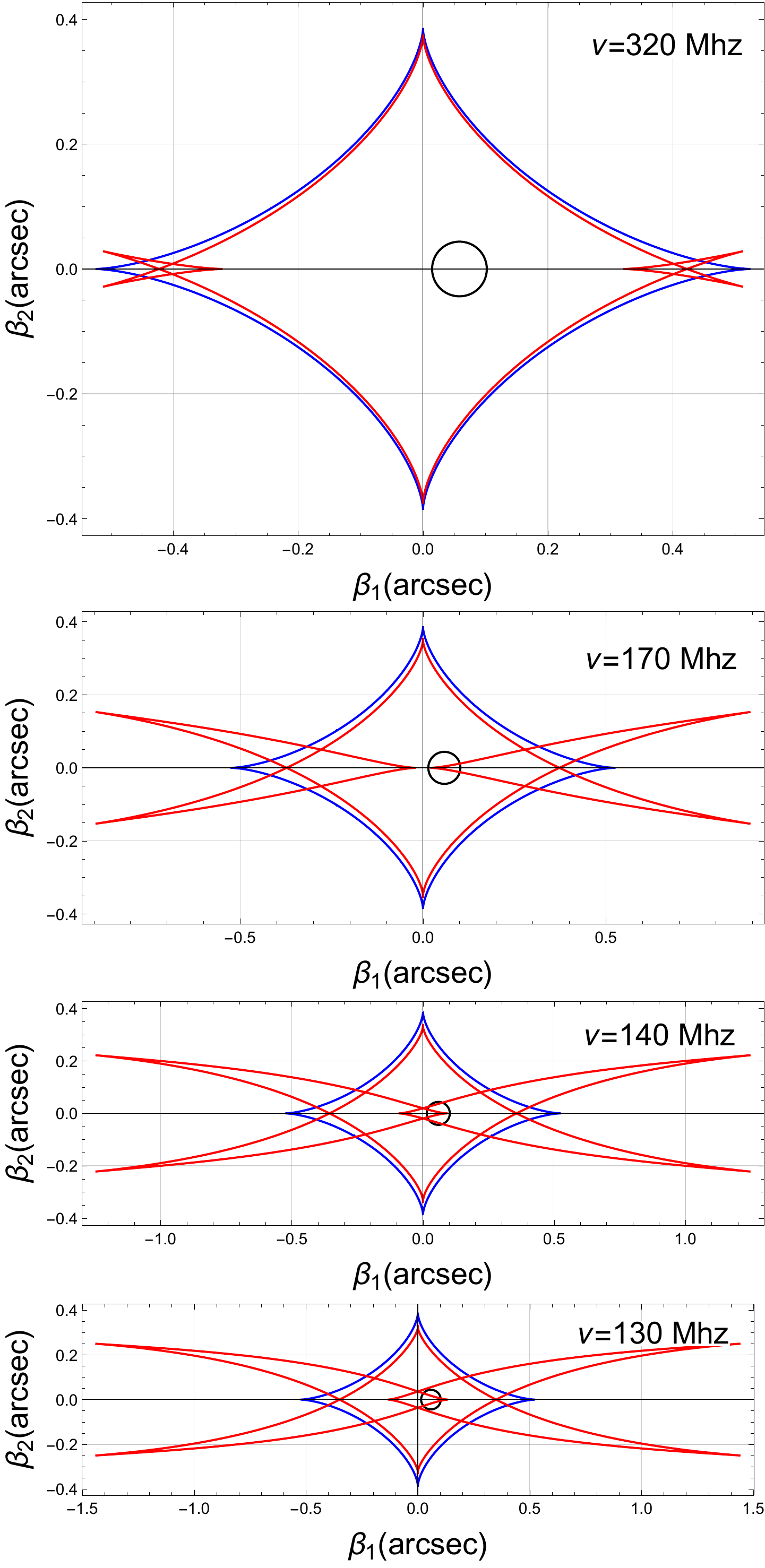}
    \caption{Caustic curves associated to the SIE 
    model of Fig. \ref{hhh2}, for a plasma disk 
    with Gaussian decay in the z direction with 
    parameters $n_0=40$cm$^{-3}$, $r_p=10$kpc, 
    $z_0=1$kpc, $\sigma_c=180$km/s , $\eta=0.3$ 
    for a circular source with radius 
    $\delta\beta_s=0.06\theta_E$ centered at 
    $\delta\beta_{10}=0.08\theta_E$, 
    $\delta\beta_{20}=0. 0$. Blue and red lines 
    correspond to caustic curves in the pure 
    gravity case and in the plasma case, 
    respectively. In addition the position of the 
    source is shown in black line. 
    }
\label{caustics}
\end{figure}

In Fig. \ref{caustics} we can see the caustic curves associated with Fig. \ref{hhh2}. In blue we see the caustic curves for the pure gravity case in the form of an astroid while in red the caustic curves for the case with plasma are shown for different observation frequencies. In black we can see the relative position of the source with respect to these curves. The effect that the plasma has for lower frequencies is notable, which is correlated with the images that are formed in the lens plane. For an observation frequency of 320Mhz we see that although the effect of the plasma on the caustic curves is appreciable, the images change very little because the source is relatively far from them. However, for lower and lower frequencies we see substantial effects in the images formed along the axis $z=0$ in the lens plane, a situation that is consistent with an increasing approximation between the source and the caustic curves. For an observation frequency of 130Mhz we can even see that the plasma produces a change in the multiplicity of the images that is perfectly distinguishable, coinciding with an overlap between the source and the caustic curves in multiple places. In particular, at a frequency of 130MHz, the source is located within the left-hand cusp, which appears near the center, without having fully entered the interior of the right-hand cusp nearby. This explains why the image inside the Einstein ring is split into three images on the positive abscissa axis in the rightmost figure of Fig. \ref{hhh2}, in accordance with the general theory of image formation near caustic curves (see \cite{Schneider-book} and references therein). However, the same is not observed in the image on the left, which still retains the shape of an elongated arc (also in agreement with the general theory). It should be noted that, since we are dealing with singular gravitational potentials, the odd number images theorem does not apply to the situations studied here.

Finally, in Fig. \ref{several-images} we see simultaneously the images that are formed (sub-figure below) for different relative positions of the source with respect to the caustic curves (sub-figure above). In this way, it is possible to better appreciate the importance of caustic curves in the morphology and multiplicity of images in the lens plane.

\begin{figure}
	\includegraphics[width=\columnwidth]{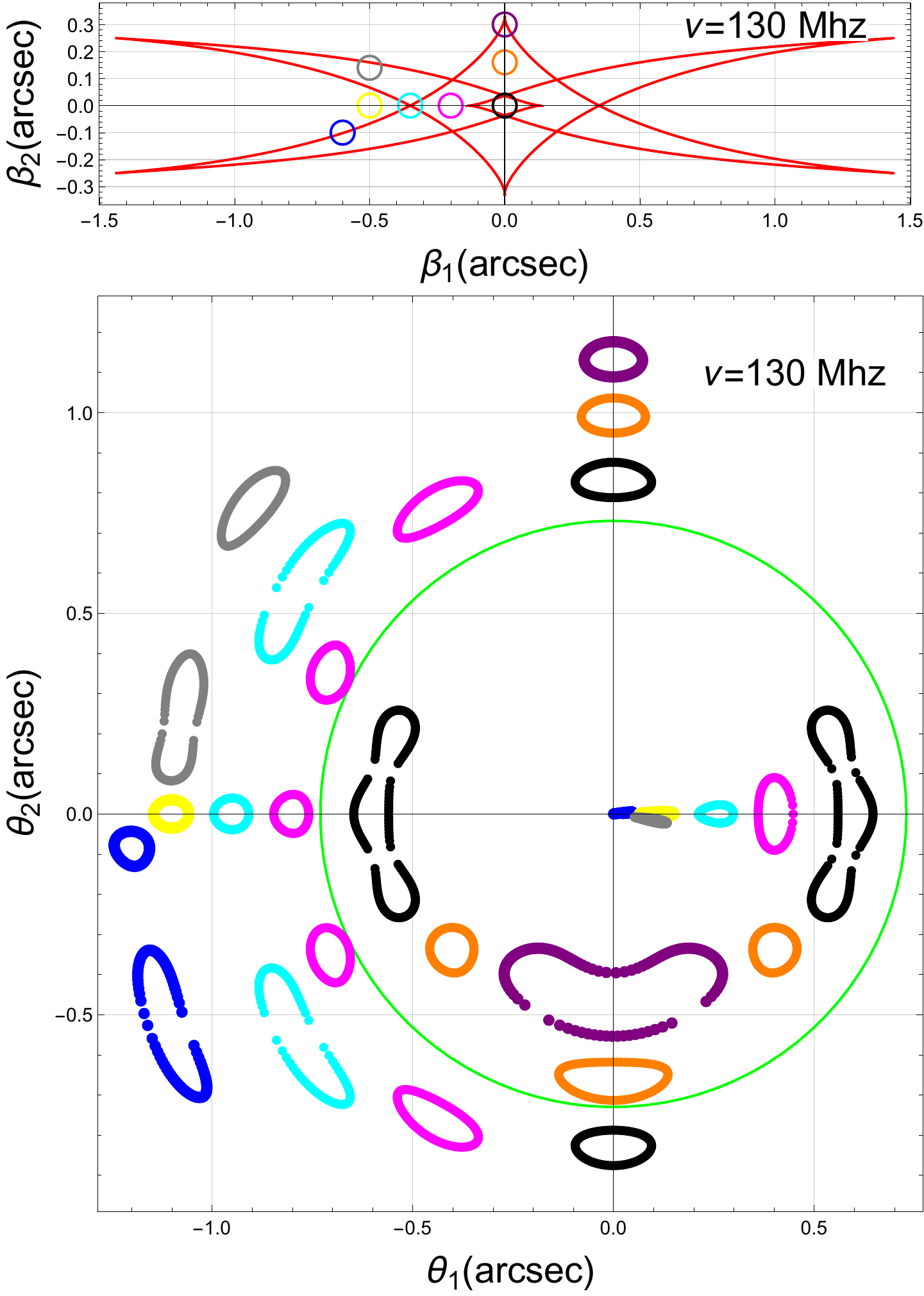}
    \caption{Images for different positions of the source with respect to the caustic curves for the same configuration of \ref{hhh2} with observation frequency $\nu=130$Mhz.}
\label{several-images}
\end{figure}

\section{A 3D spheroidal model}
\label{3D-model}
\subsection{The model}
It is our main interest in this section to put in perspective the two dimensional simple lens models 
discussed previously by comparing them with enhanced models coming from a 
volumetric distributions.
There are a couple reasons that motivate this further analysis, let us mention for example that even 
though these 2D models are useful for fitting and quantifying many astrophysical systems it is often
the case that spheroidal symmetric distributions do not project to surface mass densities with ellipsoidal 
symmetry; as in the case of SIEs.
Another usual feature of elliptical models is that they are based on two dimensional isodensity or 
equipotential curves with ellipses having the same eccentricity, instead in some case it could be 
desirable to have at hand models that tends to be spherically symmetric in the limit or large distance 
from the center of the distribution; this is not the case for the models previously studied.
For instance, the spheroidal symmetries appear naturally when one considers generalized mass distribution 
in Newtonian gravity which leads for example to the so-called third Newton theorem \citep{Binney2008}.

Below we present the description of static and spheroidal lenses with an approach that has some 
differences with respect to the usual treatments. 
Here, we will prefer to deal with a geometrical model describing the spacetime associated to the lens 
and built an spheroidal symmetric model based on the standard oblate spheroidal 
coordinates\citep{moon1988field}, where the spheroids of symmetry are confocal; this means that when 
the radial coordinate associated to spheroids growth its eccentricity $\epsilon$ decreases.
So, below we present a geometry that has oblate symmetry and becomes a direct generalization of the 
well known singular isothermal profile in spherical symmetry to the spheroidal case.
The model is determined by the metric

\begin{equation}\label{eq:disc-case-lin-final}
\begin{split}
ds^2 
=& 
\bigg( \frac{r}{r_0}\bigg)^{4 \sigma^2}
dt^2 
-  
\bigg(
\frac{\Sigma}{r^2 + r_\mu^2   }
+ \frac{2 M(r)}{r}
\bigg)
dr^2 \\ 
&- \Sigma d\theta^2 
- \left( r^2 + r_\mu^2 \right) \sin^2(\theta)
d\phi^2
;
\end{split}
\end{equation}
where 
\begin{equation}
M(r) = 2 \sigma^2 r,
\end{equation}
and $r_0$ and $r_\mu$ are fixed parameters.
Note that $\sigma$ here does not have any unit ($\sigma \equiv \sigma_c / c$).
If $r_\mu=0$ the spacetime becomes spherically symmetric and would have a mass-energy 
density of isothermal form given by
\begin{equation}\label{eq:rho+3d+model}
\rho_\text{sphe}(r) 
= \frac{1}{2 \pi r^2}\frac{\sigma^2}{1 + 4 \sigma^2}
\simeq \frac{\sigma^2}{2 \pi r^2};
\end{equation}
while its spacelike matter components would be $P_r(r) = 0$ and 
$P_{\theta}(r) = \sigma^2 \rho_\text{sphe}(r)$ (and so $ P_{\theta}(r) \ll \rho_\text{sphe}(r)$ 
since typically $\sigma^2 \ll 1$).
That is, when $r_\mu =0$ the above metric is a peculiar singular isothermal profile;
its mass-energy density is only determined by the mass function $M(r)$ while the 
timelike component of the metric ensures a model with negligible pressures and 
vanishing stresses.
The behaviour of the solution in the limit $r_\mu = 0$ was in fact one of the criteria
adopted to find the line element \eqref{eq:disc-case-lin-final} since in this case, this
lens will display similar phenomenology than the spherical analogues of the SIE model 
of the previous sections.
Instead, if $\sigma = 0$ one obtains the flat line element in the standard oblate 
spheroidal coordinates.
The eccentricity of the oblate spheroids of $r=$constant is given by $\epsilon^2 = \frac{r_\mu^2}{r^2+ r_\mu^2}$.
As we previously mentioned, this becomes a subtle difference with respect to the SIE 
model where the ellipses of symmetry have a constant eccentricity, namely 
$\epsilon_{_{SIE}}^2 = \frac{2\eta}{1 + \eta} =$constant.

Then, in order to compare both models a suitable choice of the parameter $r_\mu$ 
has to be made to fit a SIE characterized by $\eta$=constant.

For $r_\mu \neq 0$, the model presented in this section is a kind of natural generalization 
to oblate spheroidal symmetry of the singular isothermal profile, in particular it is interesting 
to note that even the stresses given by the component $T_{r \theta}$ of the energy momentum tensor
are present, they become negligible.

The following sections show the behaviour of this lensing in both weak and strong regimes
when a plasma is present or absent.

\subsection{Weak lensing optical scalars}\label{sec:cosm}
\subsubsection{Achromatic lensing}
In this subsection we will present the calculation of the optical scalars associated to the model 
of the previous section. 
%
The methods used in this section differ from those introduced in section \ref{sec:Basic_equations};
to properly account the full geometry of the lens we will base our computation on the framework presented 
in references\cite{Gallo11, Boero:2018a, Crisnejo:2019xtp}.
This framework allows to deal with the full contributions coming from the curvature of the lens, in 
particular it takes into account for the possible spacelike contributions of the energy momentum tensor 
which are usually neglected in most methods based on linearized gravity \citep{Schneider92}.
We will use the following expression valid when the cosmological background is explicitly taken into account
\citep{Boero:2018a}:
\begin{align}
\kappa =& \left( 1 - \kappa_c \right)\kappa_L + \kappa_c, \label{eq:kappa}
\\
\gamma =& \left( 1 - \kappa_c \right)\gamma_L
\label{eq:gamma}
\end{align}
where now the cosmological contribution $\kappa_c$ to the expansion is given by
\begin{equation}
\kappa_c = 1 - \frac{D_A(\lambda)}{\lambda},
\label{eq:kappa_C}
\end{equation}
and where the lens intrinsic contribution to the expansion and the shear are given by 
\begin{align}
\kappa_L =& \frac{D_{A_l} D_{A_{ls}}}{D_{A_s}} \int_{\lambda_o}^{\lambda_s} \Phi_{00} d\lambda, 
\label{eq:kappaL}
\\
\gamma_L =& \frac{D_{A_l} D_{A_{ls}}}{D_{A_s}} \int_{\lambda_o}^{\lambda_s} \Psi_{0} d\lambda;
\label{eq:gammaL}
\end{align}
where $\lambda$ and $D_A$ denotes the geometric affine distances and angular diameter distances
respectively, and subindices $_o$ and $_s$ refers to values at the observer and the source
respectively. The quantities $\Phi_{00} ~=~ -\frac{1}{2} R_{ab} \ell^a \ell^b$ and 
$\Psi_{0} ~=~ C_{abcd} \ell^a m^b \ell^c m^d$ are the Ricci curvature scalar 
and Weyl curvature scalars of the GHP formalism \citep{Geroch:1973am} with respect to a null 
tetrad  $(\ell^a, m^a, \bar{m}^a, n^a)$ adapted to the path of the photon under consideration.
In weak lensing regime only first order effects on the curvature are relevant and so 
in this case the null geodesic and the null tetrad are taken to be those corresponding to 
unperturbed null geodesic of the background (as described for example in \cite{Gallo11}). 
Under such approximation, the limits of the integral range from $\lambda_o$ to $\lambda_s$ which are 
the values of the affine parameters at the observer and at the source respectively.

For first order effects on the curvature as those present in weak lensing regime,
In particular, for first order effects on the curvature, the null geodesic as well 
as the null tetrad are taken to be those corresponding to unperturbed null geodesic 
of the background (as described for example in \cite{Gallo11}) and the limits of the 
integral range from $\lambda_o$ to $\lambda_s$ which are the values of the affine
parameters at the observer and at the source respectively.

\subsubsection{Chromatic lensing}
When a static plasma is present on the static geometric lens, similar expressions to the 
previous sub-subsection hold; in such case, the curvature scalars are those associated to 
the \textit{Gordon-like optical metric} introduced in \cite{Crisnejo:2019xtp}.
That metric has the property that the projected spacelike orbits of massless particles
on the surfaces $t=$constant coincides with those of the spacetime metric; and,
since deflection angles are essentially deduced from the spacelike orbits, the expressions
for the bending angle and optical scalar in a medium filled with a plasma are
identical to those of the spacetime metric. 
Then, one can show that similar expressions to those of equations \eqref{eq:kappa},
\eqref{eq:gamma}, \eqref{eq:kappaL} and \eqref{eq:gammaL} remains valid if one 
replace the curvature scalar $\Phi_{00}$ and $\Psi_0$ of the spacetime metric by 
its analogues computed in the Gordon-like optical metric; this is:
$\Phi_{00} \to \Phi_{00_G}$ and $\Psi_0 \to \Psi_{0_G}$, where the subindex $G$ stems 
for Gordon-like line element.
The associated Gordon-like optical metric to equation \eqref{eq:disc-case-lin-final}
for a static plasma with spheroidal symmetry is
\begin{equation}
\begin{split}
ds_{G}^2 
=& 
\frac{1}{n(r)^2}\bigg( \frac{r}{r_0}\bigg)^{4 \sigma^2}
dt^2 
-  
\bigg(
\frac{\Sigma}{r^2 + r_\mu^2}
+ \frac{2 M(r)}{r}
\bigg)
dr^2 \\ 
&- \Sigma d\theta^2 
- \left( r^2 + r_\mu^2 \right) \sin^2(\theta)
d\phi^2
; \label{eq:disc-case-lin-final+plasma}
\end{split} 
\end{equation}
where $n(r)$ denotes the refractive index of the ionized medium.

\subsubsection{Including a plasma model: the refractive index}\label{subsubsec:Includ_Plasma}

For our spheroidal symmetric lenses we will use a plasma distribution that is
consistent with that symmetry, for this purpose we will consider a plasma density of the form
\begin{equation}\label{eq:elect+numb+dens+3dModel}
n_e(r) = n_p e^{-\frac{r}{r_p} } 
,
\end{equation}
with $r_p$ and $n_p$ both constants. 
Here it is important to recall that $\omega$ is the angular frequency that one would measure at 
the location of the plasma, so that if we consider an observing frequency $\nu$ then we will have
\begin{equation}
\omega = \frac{2\pi \nu (1 + z_l)}{\sqrt{g_{tt}}},
\end{equation} 
with $g_{tt}$ the timelike component of the line element \eqref{eq:disc-case-lin-final}.

\subsubsection{Numerical results}\label{subsubsec:Numer_resul_weakl}
\begin{figure*}
\centering
\includegraphics[width=\columnwidth]{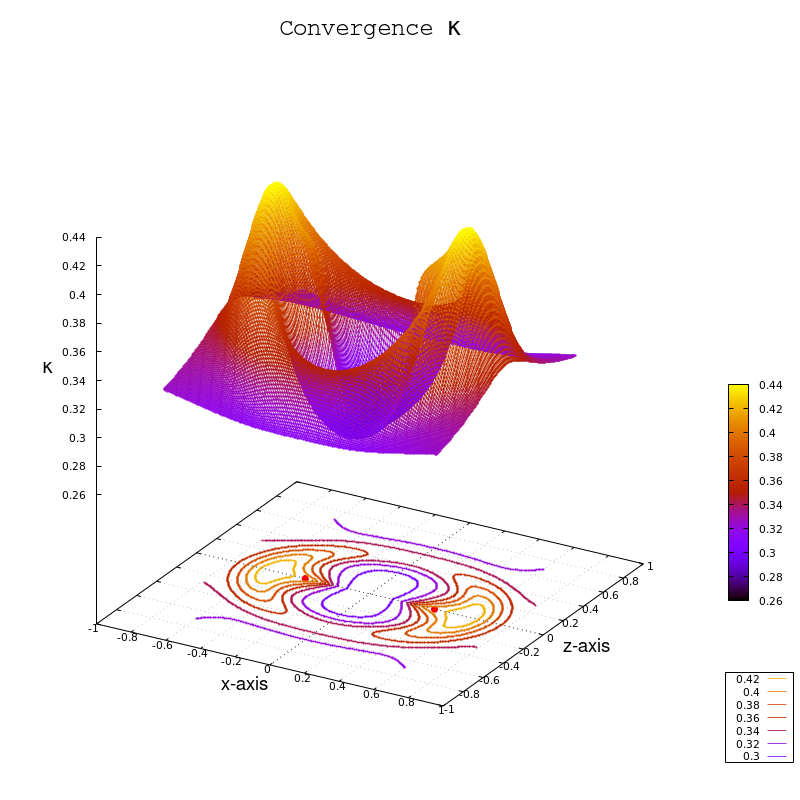}
\includegraphics[width=\columnwidth]{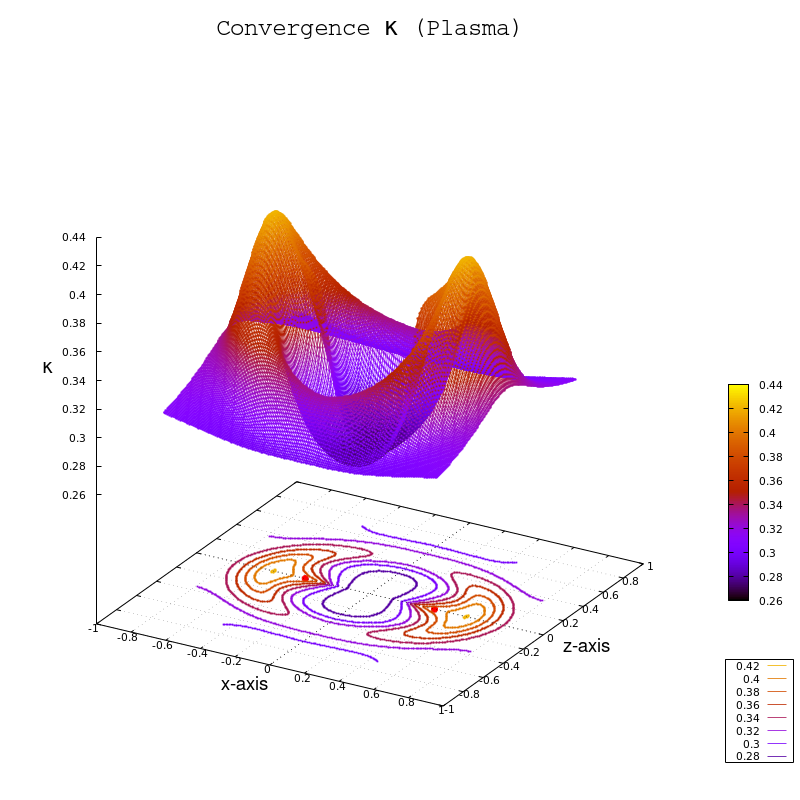}
\caption{Convergence for a geometry with $r_\mu = 0.7 r_E$, $\sigma_c = 180\text{km}\text{s}^{-1}$ and $r_0 = 0.32\text{Mpc}$.
The plasma model is show next to the pure gravity case and has been chosen such that: $\nu_o = 180\text{MHz}$, $n_p =60\text{cm}^{-3}$, $r_p = 10\text{kpc}$.
Red dots in the figures indicates the position of the foci that corresponds to the ellipse generating the oblate ellipsoid.}
\label{fig:kappagamma_r_mu-0_7rE}
\end{figure*}

\begin{figure*}
\centering
\includegraphics[width=\columnwidth]{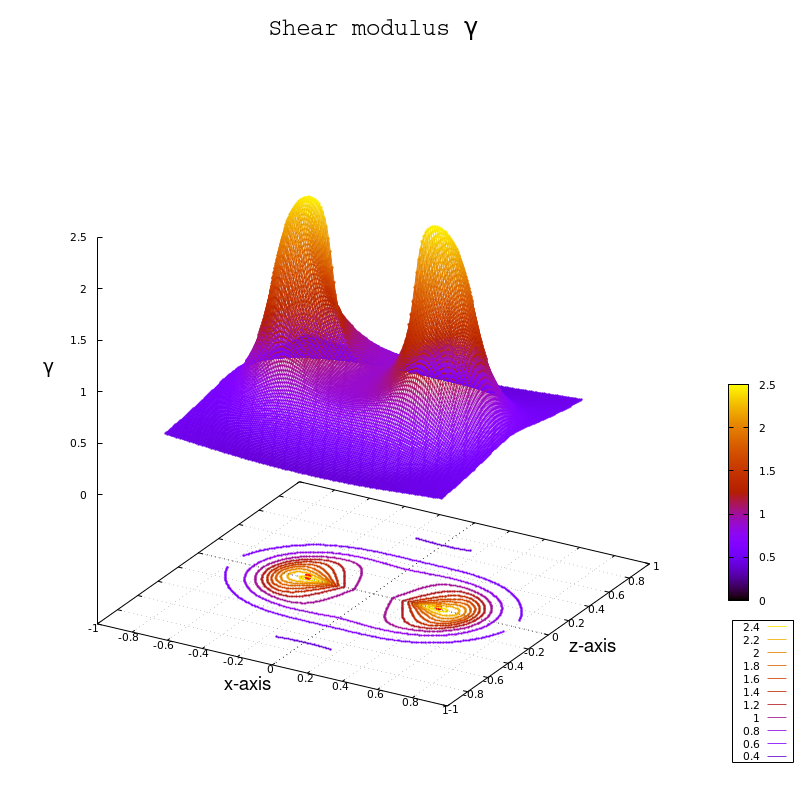}
\includegraphics[width=\columnwidth]{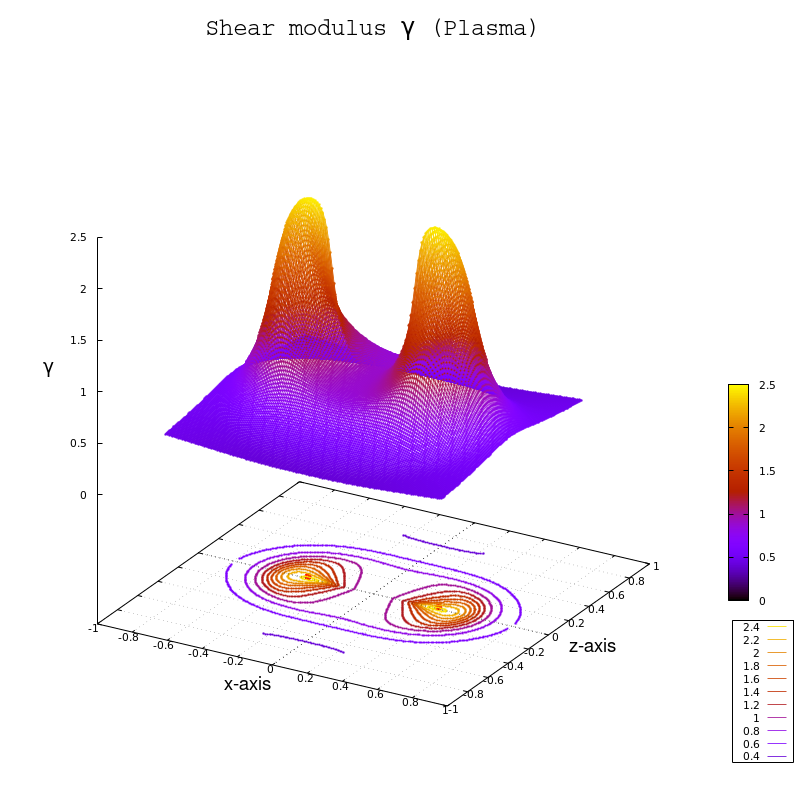}
\caption{Shear for a geometry with $r_\mu = 0.7 r_E$, $\sigma_c = 180\text{km}\text{s}^{-1}$ and $r_0 = 0.32\text{Mpc}$ plus a plasma model with the following parameters: $\nu_o = 180\text{MHz}$, $n_p =60\text{cm}^{-3}$, $r_p = 10\text{kpc}$.
Red dots in the figures indicates the position of the foci that corresponds to the ellipse generating the oblate ellipsoid.}
\label{fig:gamma_r_mu-0_7rE}
\end{figure*}

\begin{figure*}
\centering
\includegraphics[width=\columnwidth]{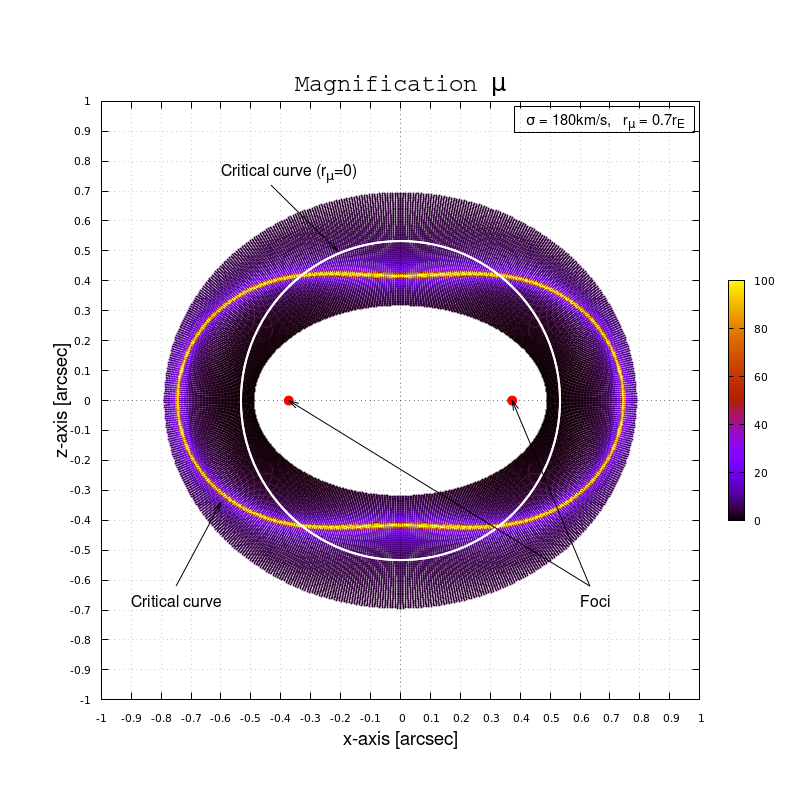}
\includegraphics[width=\columnwidth]{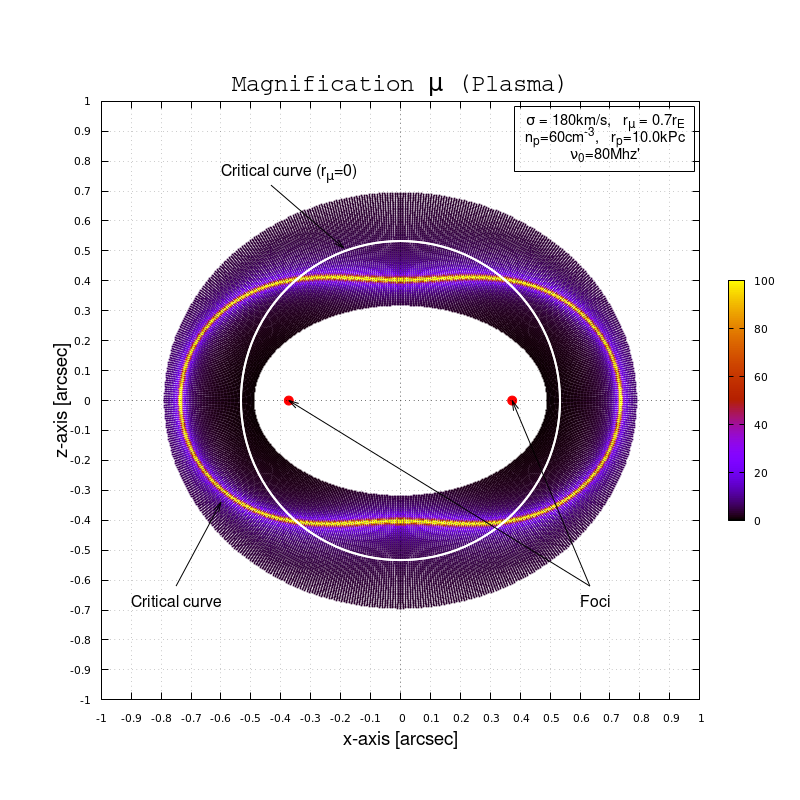}
\caption{Magnification factor for a geometry with $r_\mu = 0.7 r_E$, $\sigma_c = 180\text{km}\text{s}^{-1}$ and $r_0 = 0.32\text{Mpc}$.
The plasma model is show next to the pure gravity case and has been chosen such that: $\nu_o = 180\text{MHz}$, $n_p =60\text{cm}^{-3}$, $r_p = 10\text{kpc}$.
Red dots in the figures indicates the position of the foci that corresponds to the ellipse generating the oblate ellipsoid.
For reference with the spherically symmetric case ($r_\mu = 0$) we have also included in the form of a white ring the Einstein ring for the case of pure gravity.}
\label{fig:r_mu-0_7rE}
\end{figure*}

With the aim to illustrate the typical behaviour of the oblate spheroidal geometry 
we show below the result of a numerical computation of the optical scalar in Figs. 
\ref{fig:kappagamma_r_mu-0_7rE} and \ref{fig:gamma_r_mu-0_7rE}, 
and the magnification of a lens with such geometry in Fig. \ref{fig:r_mu-0_7rE}. 
Left panels of these figures correspond to pure gravity while right panels to the spheroidal
geometry with a plasma.  	
The parameter of the geometry in this example are $r_\mu = 0.7 r_E$, 
$\sigma_c = 180 \mathrm{km s}^{-1}$; where the radial scale $r_E$ corresponds to the Einstein's radius 
of a spherically symmetric singular isothermal profile, this is 
$r_E = 4\pi \sigma_c^2 \frac{D_l D_{ls}}{D_s}$.
For the plasma we choose $n_p = 60 \mathrm{cm}^{-3}$ and $r_p = 10 \mathrm{kpc}$.
We have used the affine distances 
$\lambda_l = 169.55$Mpc and 
$\lambda_s = 395.164$Mpc; which, when expressed as angular diameter distances 
are $D_{A_l} = 169.528$Mpc and $D_{A_s} = 394.856$Mpc respectively on the flat ($k=0$) Friedman-Robertson-Walker model that we have chosen (see introductory section).

The two figures Figs. \ref{fig:kappagamma_r_mu-0_7rE} and \ref{fig:gamma_r_mu-0_7rE} show the 
projection of $\kappa$, $\gamma$ in the plane $(x,z)$ of the lens together with their level sets.
They exhibit a non-trivial structure near of the foci of the spheroids associated with the geometry
and only far from the foci the level set resemble at some extent to elliptical shapes.
Fig. \ref{fig:r_mu-0_7rE} show the position of the critical curves (sharp yellow contour) that would 
correspond to deformations of the Einstein ring of the case $r_\mu = 0$.

The examples correspond to the simplest observational setting were a distant observer is located
at $\theta = \frac{\pi}{2}$ and where the spheroidal geometry appear to her/him not tilted.
The case of general orientations can be handled by and appropriated rotation of the spheroids 
local frame.
	
The numerical integration of equations \eqref{eq:kappa} and \eqref{eq:gamma} in all the situation
was accomplished by the use of a Gauss-Legendre quadrature of order 7 while the computation of 
the curvature scalar was done with the use 
of xAct suite\citep{xact}.


\subsection{Relation with elliptical models}
In order to attempt a comparison between the geometric model and the SIE model, an appropriated 
choice of parameters has to be made.
For example, one could try to fit critical curves of one model to the other or to associate the
elliptical shape of the SIE potential with an appropriated projected ellipsoids among others possibilities.
For this study we combine this two criteria to fit similarities between features of both models.
We note first that the equipotential curves of $\psi_\text{grav} = k$, 
\begin{equation}
\theta_k = \frac{k}{\theta_E  \sqrt{1 - \eta \cos(2\phi)}},
\end{equation}
define ellipses with semi-major axis $a$ along $\theta_1$ and semi-minor axis $b$ along
$\theta_2$, in particular one has that:
\begin{align}
1 - \eta =& \frac{k^2}{\theta_E^2 a^2}, 
\\
1 + \eta =& \frac{k^2}{\theta_E^2 b^2}.
\end{align}
Dividing both expressions and rearranging terms to isolate $\eta$ one obtains
\begin{equation}
\eta = \frac{a^2 - b^2}{a^2 + b^2},
\end{equation}
which gives a relation between the parameter $\eta$ only in terms of the principal axis of the ellipses.
Let us note that if one takes $k = \chi \theta_E^2$ then $\theta_{k_0} = \chi \theta_E / \sqrt{1 - \eta}$ 
and $\theta_{k_{\pi/2}} = \chi \theta_E / \sqrt{1 + \eta}$.
Similarly than before, one can see that:
\begin{equation}
\eta = \frac{\theta_{k_0}^2 - \theta_{k_{\pi/2}}^2}{\theta_{k_0}^2 + \theta_{k_{\pi/2}}^2}.
\end{equation}

On the other hand, in the spheroidal geometry one has that $a^2 = r^2 + r_\mu^2$ and $b^2 = r^2$,
so that one has
\begin{equation}
\begin{split}
r_\mu^2 = a^2 - b^2 =&  D_l^2\left( \theta_{k_0}^2 - \theta_{k_{\pi/2}}^2 \right)
\\
=& \chi^2 D_l^2 \theta_E^2 \frac{2\eta}{1 - \eta^2}
\equiv \chi^2 r_E^2 \frac{2\eta }{1 - \eta^2},
\end{split}
\end{equation}
where we have defined
\begin{equation}
r_E \equiv D_l \theta_E = 4 \pi \frac{\sigma_c^2}{c^2}  \frac{D_l D_{ls}}{D_s}.
\end{equation}

Preliminary comparison of the critical curves of both models suggests to consider
$\frac{1}{\chi}=1.16$. In terms of the relation $\eta = \eta(r_\mu)$ one obtains 
for example $\eta(0.3 r_E)=0.06033$, $\eta(0.4 r_E)=0.10643$, 
$\eta(0.5 r_E)=0.16369$, $\eta(0.6 r_E)=0.22945$, $\eta(0.7 r_E)=0.30000$.

\subsection{Strong lensing of a disklike oblate spheroidal geometry}
Let us consider now large distortions of background sources due to our geometric model; 
such a strong lensing effects will be associated to sources near to caustic points.
It is then necessary to determine the region of caustics for the model and afterwards 
to look at the arcs produced by a small circular source near to the caustic.
Let us to realize that in the full geometric treatment we do not have, in general, 
an explicit lens equation in terms of bending angles, so that the image generation as well as
the location of caustics have to be performed by means of the ray tracing of the null geodesics of 
the geometry \eqref{eq:disc-case-lin-final} in the case of pure gravity, or null geodesics of 
equation \eqref{eq:disc-case-lin-final+plasma} if we take into account the chromatic effect 
of a static plasma.
The numerical solution requires the integration of the exact null deviation geodesic equations 
and null geodesic equations in order to determine the caustics and images that we present
below.

\subsubsection{Numerical implementation}
Our ray-tracing code uses a classical Runge-Kutta integrator pair of order 7-8 from the suite 
RKSuite\citep{rksuite-90} with high accuracy to find the solution to the path of the photons 
from the observer position. 
The code additionally computes the solution of the geodesic deviation equation and therefore 
allows us to compute the optical scalars along the central null geodesics of a thin bundle. 
In particular, in order to asses the position of caustics and critical curves we integrate the
equations to find those points where the magnification factor becomes divergent.
The search of the angular directions that corresponds to points comprising caustic curves or the 
points corresponding to the source is implemented through iterative approximations. 
In the sky of the observer one starts with two initial guesses taken along a radial direction 
in the plane perpendicular to the line of sight and in a next step we run the mid-point
algorithm to correct the initial geodesics to the closer ones to the specified target,
the position of the source or the criteria that the geodesic reach a divergence beyond 
certain threshold.
For instance, in our examples below, the typical size of the images have angular size of order 
$0.1-1$arcsec; for them we have taken a error tolerance in the determination smaller than 
$10^{-4}$arcsec. In the case of caustic, we have required that geodesics having magnifications
factor greater than $10^6$ should be considered as effectively passing through a caustic point.

\subsubsection{Results}
\begin{figure}
\centering
\includegraphics[width=0.5\textwidth]{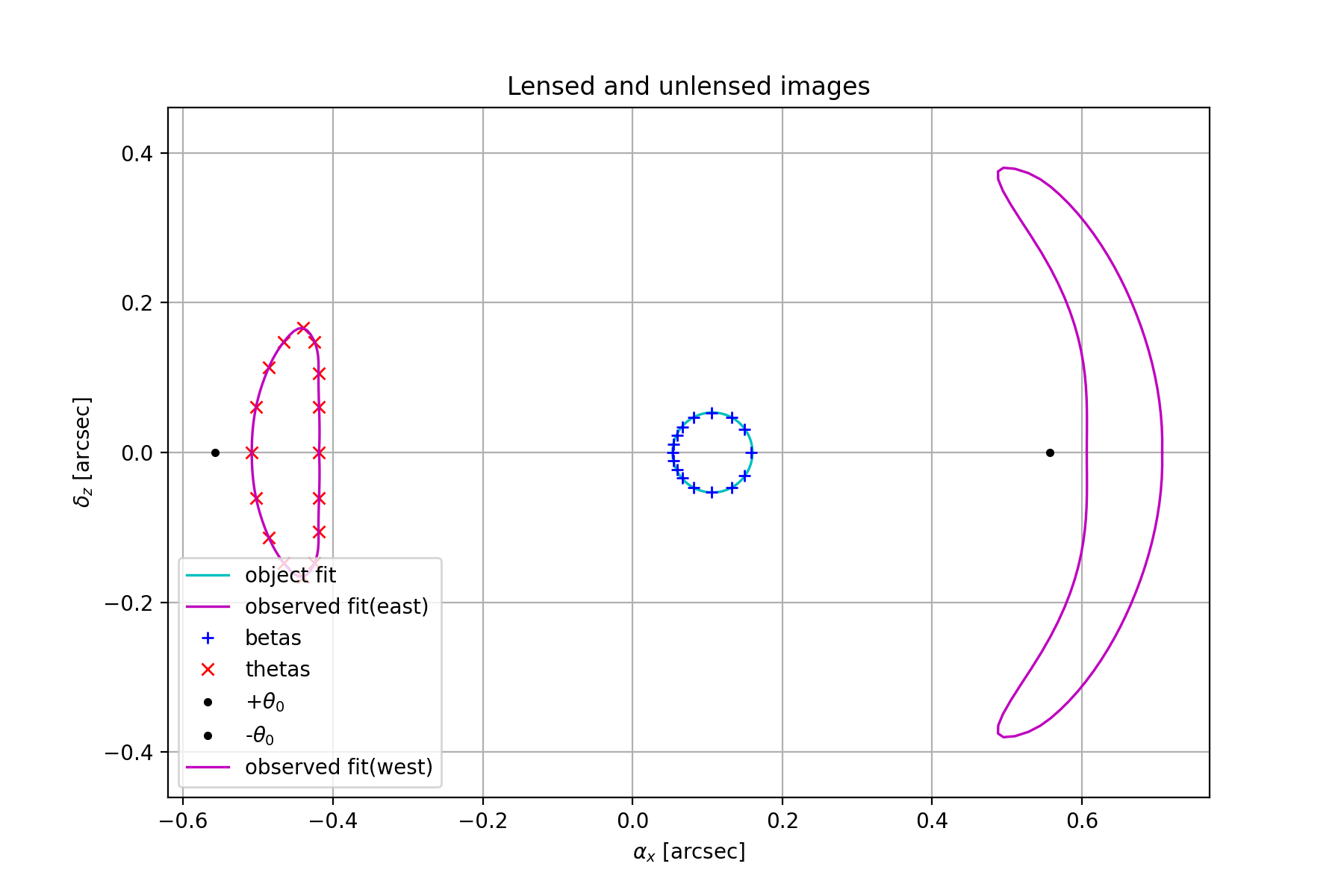}
\caption{
Pure gravity: lensed and unlensed images of a circular source intersecting the 
caustic region of the model characterized by $r_\mu = 0.3r_E$, $\sigma_c = 180\text{km}\text{s}^{-1}$ and 
$r_0 = 0.32\text{Mpc}$.
Blue marks indicate those points of the source employed to build the 
images of the arcs; in red the small crosses indicate the observed
position of those points in the source.
The black points on the horizontal axis indicates the angular position of the 
Einstein ring in the spherical case.
}
\label{fig:circulos}
\end{figure}

\begin{figure}
\centering
\includegraphics[width=0.5\textwidth]{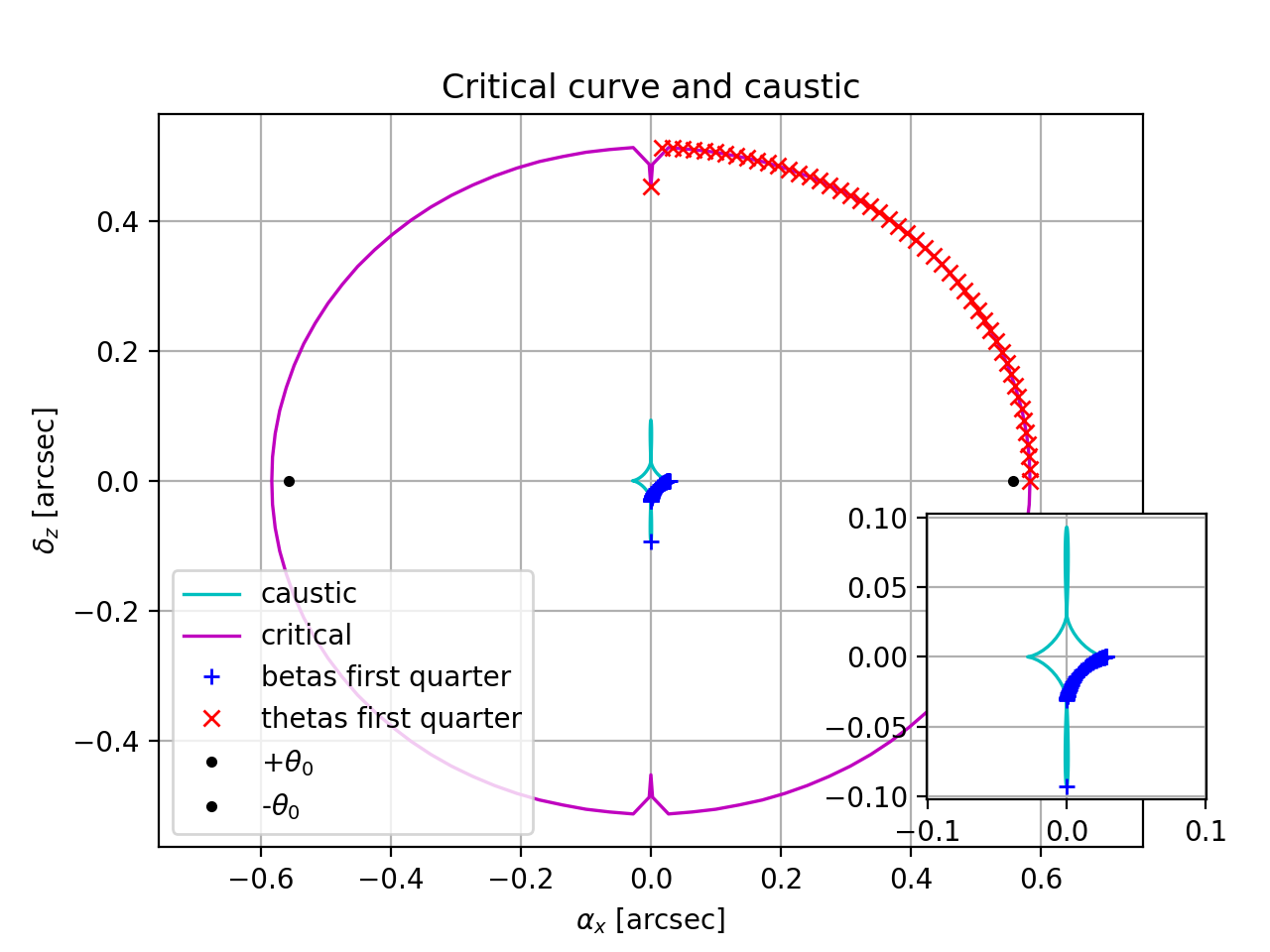}
\caption{
Pure gravity: critical and caustic curves for an oblate spheroidal model with a small deviation
from spherical symmetry. 
The geometry is characterized by $r_\mu = 0.3r_E$, $\sigma_c = 180\text{km}\text{s}^{-1}$ and 
$r_0 = 0.32\text{Mpc}$.
Blue marks shown on the caustic are in correspondence with the red marks in the 
critical curve. 
The black points on the horizontal axis indicates the angular position of the 
Einstein ring in the spherical case.
}
\label{fig:critical+caustic}
\end{figure}

We present below the results of calculations in two different cases: the first one corresponds 
to a model with less separation between foci than the previous one discussed in the weak lensing 
analysis (section \eqref{subsubsec:Numer_resul_weakl}); we have taken in this case 
$r_\mu = 0.3 r_E$ while retaining the other lens parameters and distances unchanged.
In order to visualize large distortions like arcs we also placed a circular source at 
$\left( \beta_x, \beta_z \right) = \left( 0.2 \theta_E, 0 \right)$arcsec, with radius
$r_{\text{source}}$ of size $r_{\text{source}} = 0.1$arcsec.
The results in are shown in Figs. \ref{fig:circulos} and \ref{fig:critical+caustic}
for the case of pure gravity.
It is observed in the former one, the formation of large arcs at both side of the source;
one of them lies outside of the region delimited by the critical curve while the other
is inside.
The presence of two arcs is associated with the fact that the source intersect the 
caustic; this can be appreciated in Fig. \ref{fig:critical+caustic} where it is shown
the appearance of the critical curve and the caustic giving origin to it.
It is worthwhile to mention that such behaviour is in full agreement with the expected
images that one would obtain with the well known 2D models.

\begin{figure}[H]
\centering
\includegraphics[width=0.5\textwidth]
{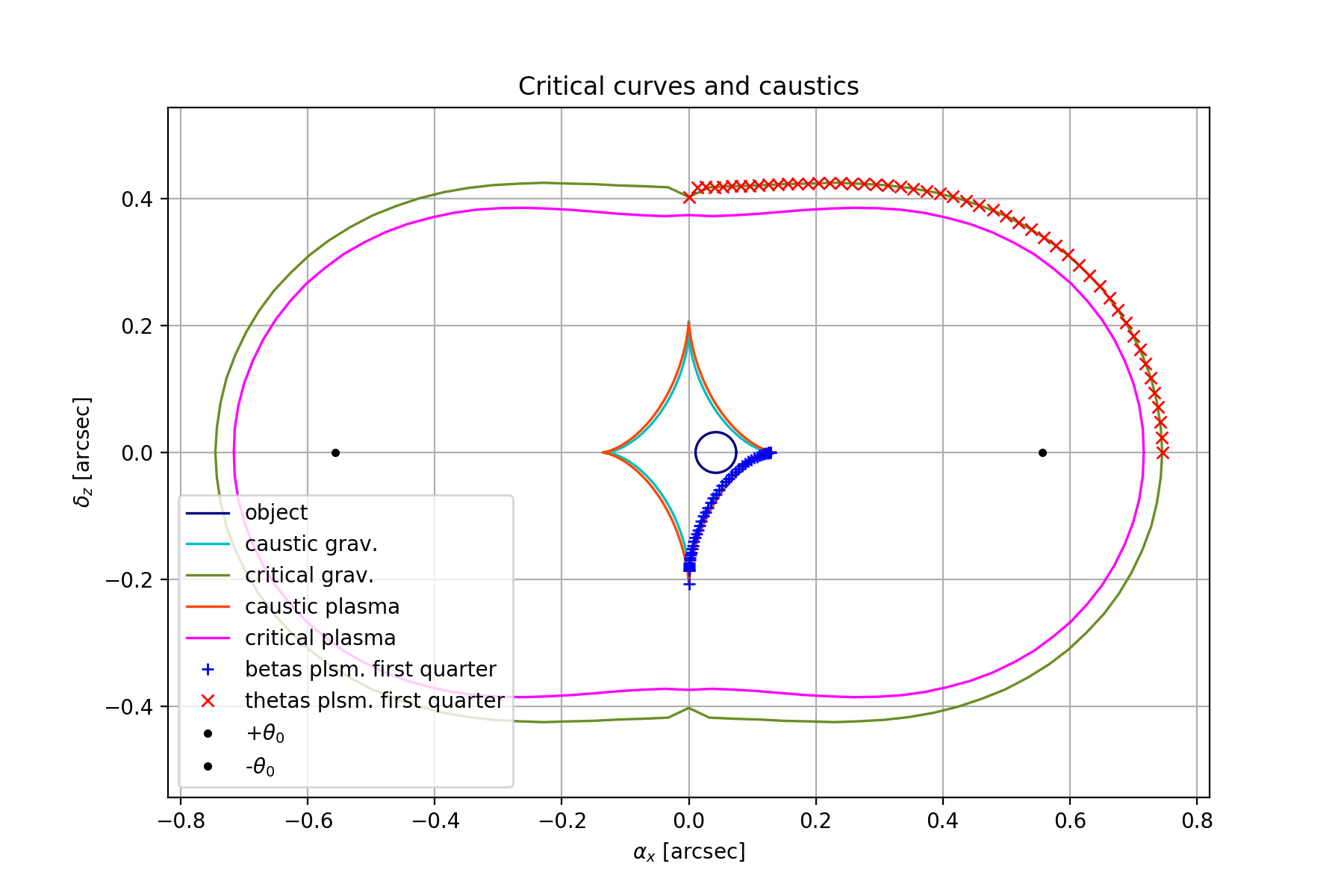}
\caption{
Critical and caustic curves for the oblate spheroidal model with and without a plasma. 
The geometry is characterized by $r_\mu = 0.7r_E$, $\sigma_c = 180\text{km}\text{s}^{-1}$ and 
$r_0 = 0.32\text{Mpc}$.
The plasma has been chosen such that $\nu_o = 80\text{MHz}$, $n_p =60\text{cm}^{-3}$, 
$r_p = 1\text{kpc}$.
The central diamond shapes in cyan and red correspond to the caustics for the 
cases of pure gravity and plasma respectively.
The correspondence between the points on the caustic (blue marks) and the 
points on the critical curve (red marks) is shown for the pure gravity case.
Black dots in the $\alpha_x$ axis are placed as reference for the position of the 
Einstein angle in the spherical case.
}
\label{fig:critical+caustic+source}
\end{figure}

\begin{figure}
\centering
\includegraphics[width=0.5\textwidth]{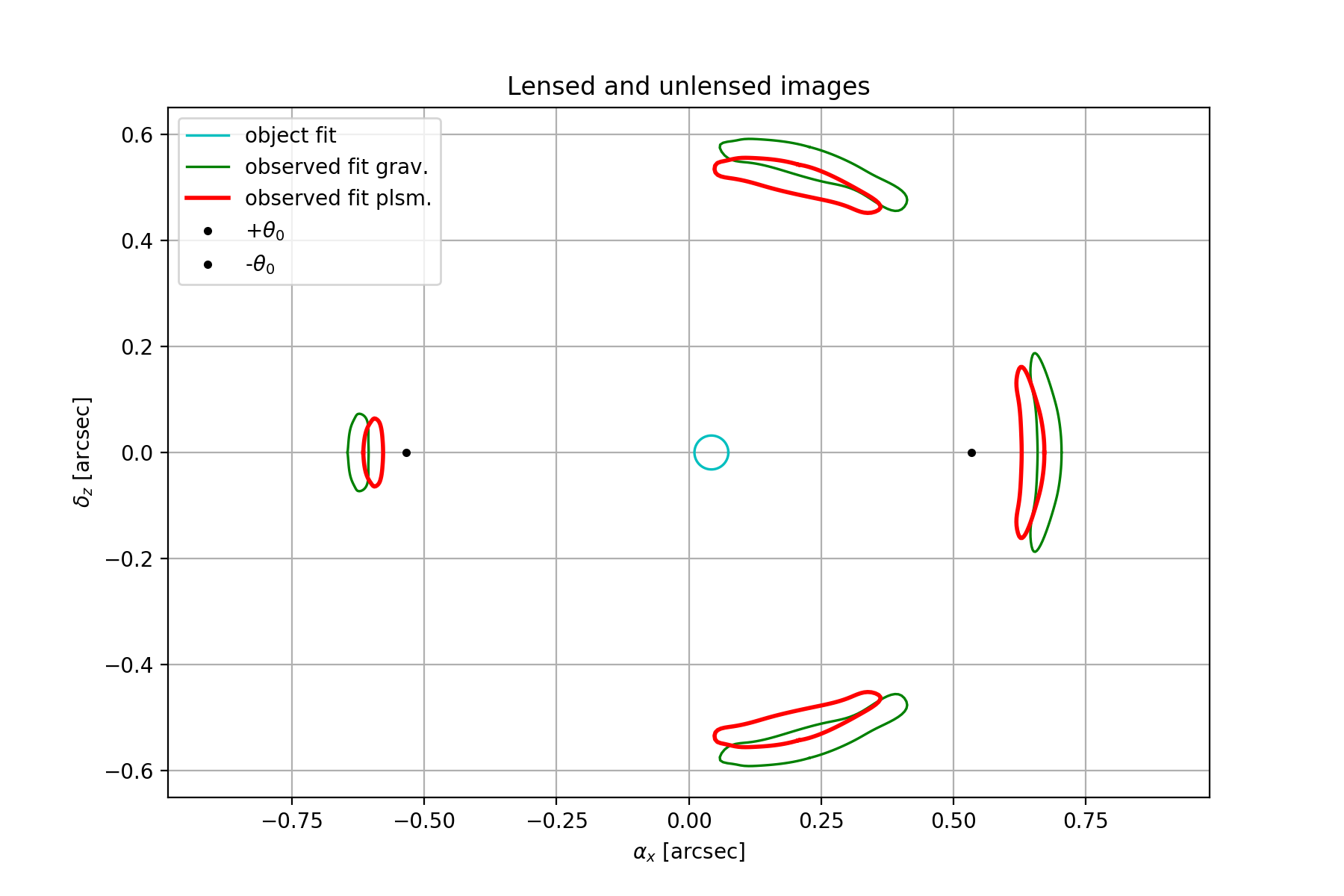}
\caption{
Lensed and unlensed images of a circular source inside of the caustic 
region shown in the previous figure, for the model the geometric model 
$r_\mu = 0.7r_E$, $\sigma_c = 180\text{km}\text{s}^{-1}$ and $r_0 = 0.32\text{Mpc}$.
The images of the circular source (cyan) are drawn in green for the pure gravity 
case.
Red curves correspond to images in the case of a plasma with $\nu_o = 80\text{MHz}$, 
$n_p =60\text{cm}^{-3}$, $r_p = 1\text{kpc}$.
}
\label{fig:imagenes_strong_plasma}
\end{figure}

We also consider for completeness, the model of section 
\eqref{subsubsec:Numer_resul_weakl} having $r_\mu = 0.7 r_E$.
The images corresponding to the circular source considered in this section are shown in Fig. 
\ref{fig:imagenes_strong_plasma} and its position with respect to the caustic (diamond shape) is 
shown in Fig. \ref{fig:critical+caustic+source}.
In both figures we have included the results when a plasma (section \ref{subsubsec:Includ_Plasma}) 
is also present.
In order to highlight the differences, in this case the plasma frequency was taken to be $\nu = 80$Mhz, 
while the remaining parameters $n_p$ and $r_p$ remaining unchanged.
Two notable changes are observed in this case with respect to the previous one: the presence of four 
images with arc shapes that appear closer to the center than those of the pure gravity case, and the 
shrink of the critical curve.
As clearly notice from Fig. \ref{fig:imagenes_strong_plasma}, the cause explaining the number of images 
should be ascribed to the position of the source in the interior of the caustic region, in agreement with 
similar results with 2D models.  
Finally, it is perhaps interesting to note that the shape of the critical curve in the pure gravity case 
fit very well the critical curve of figure \ref{fig:r_mu-0_7rE} despite both were computed with different 
methods.

%

\section{Conclusions}

In the first part of this work, after a generalization of the Alard  perturbative approach to solve the lens equations and making use of simple models for the ISM plasma around galaxies, we have shown how to include its effects on the photon propagation in the radio-frequency band to study the formation of lensed images in the strong lensing regime in an analytical perturbative way. We have described the position, shape and number of images for circular sources which are lensed by galaxies in two simple orientations, frontal and edge view. More general orientations are discussed in \cite{Tomas-tesina,Tomas-work}. For the particular cases of plasma profiles that we have used, we have shown how the number of images can increase as the observation frequency approaches the characteristic frequency of the plasma, resulting in the number of images being sensitive to the plasma profile and to the position of the source in the region of the caustic curve. 
Of course, in a real galaxy, the electronic distribution of plasma in the interstellar medium should be much more complex than in the models studied here. More sophisticated models should include the plasma concentration along arms in spiral galaxies, the bulk, disk, and also the plasma environment around satellite galaxies, etc. As previously discussed, this kind of models already exist in the literature \cite{gutierrez2010galaxy, Davies_2021, Chy_y_2018,Cordes:2002wz, Yao_2017, Yamasaki:2019htx, Price:2021gzo, van_Haarlem_2013}. However, the simple models studied here, allow us to give an idea of what kind of situations one could find when analyzing lensed systems in the low radio frequency regime, such as those expected to be observed in LOFAR+nenuFAR observatory \cite{van_Haarlem_2013,Nenufar}. It is worth noting that subarcsecond-resolution observations of bright radio sources, even at 30 MHz, have recently become possible thanks to the LOFAR observatory\cite{2022A&A...658A...9G}.
Additionaly, not only the position of the images will be affected by the presence of the plasma, but also the time delay between them, which in addition to the well-known geometric and Shapiro effects will now also depend on the plasma medium, which affects the group velocity of propagation of electromagnetic waves\cite{Bisnovatyi-Kogan:2022yzj,Er:2013efa,Er:2022lad,Tomas-tesina}. These and other studies will be developed in future works.

An interesting question is how to reconstruct the lensing properties from image observations when a gravity + plasma environment is taken into account. In such situations, parameters associated with the plasma potential model must be added to those associated with the gravitational potential. In cases where gravitational lensing can be observed in the optical regime (or where plasma effects are negligible), the gravitational potential can be separately reconstructed using conventional techniques (see, for example, \cite{meneghetti2021introduction} and references therein). However, if the source is variable, estimates of the dispersion measurement and plasma potential can be obtained by studying variations in the time delay between different images, which is chromatically dependent on the presence of plasma\cite{Wagner:2020ihx}. In cases where differences in the arrival time of signals from the images cannot be measured, the procedure becomes more complicated, but it can still be performed if the lensed system can be observed at three nearby frequencies. For more details on this method, interested readers are referred to Wagner and Er\cite{Wagner:2020ihx}.

Additionally, with the intention to further study more general models for the lens distribution, 
we have also introduced a full geometrical model which is a natural generalization of spherically
symmetric lenses with a singular isothermal profile. 
The metric proposed in this case resemble a disc-like oblate geometry that far from the center
approaches to a spherically symmetric one.
Despite the differences between this kind of models and those of section \ref{galaxy_modelling}
that do not allow for a direct comparison, we have shown that a reasonable fitting criteria can 
be attempted in order to study some qualitative features appearing in both lens models; such as
the shape of critical curves and the observed number of images that are obtained depending on the 
position of source with respect to caustics. 
While two-dimensional models are certainly useful for studying these types of problems, it is important to note that our consideration of the three-dimensional spheroidal model represents a significant step toward constructing more realistic models. Furthermore, in Section \eqref{sec:cosm}, we have taken into account the influence of the cosmological framework on these systems, further reinforcing the realism of our models.
A further analysis on the properties of these kind of model will be presented
elsewhere.

\section*{Acknowledgements}
We are very grateful to Adam Rogers for  illuminating discussions and valuable comments. The authors would like to acknowledge the valuable comments and suggestions provided by an anonymous referee, which helped to improve our paper. We acknowledge support from CONICET and SeCyT-UNC.


\begin{thebibliography}{10}

\bibitem{Narayan:1996ba}
Ramesh Narayan and Matthias Bartelmann.
\newblock {Lectures on gravitational lensing}.
\newblock In {\em {13th Jerusalem Winter School in Theoretical Physics:
  Formation of Structure in the Universe}}, 6 1996.

\bibitem{Schneider92}
P.~Schneider, J.~Ehlers, and E.E. Falco.
\newblock {\em Gravitational lenses}.
\newblock Springer-Verlag, 1992.

\bibitem{perlick2000ray}
V.~Perlick and Springer-Verlag.
\newblock {\em Ray Optics, Fermat’s Principle, and Applications to General
  Relativity}.
\newblock Lecture Notes in Physics Monographs. Springer, 2000.

\bibitem{Mao:2012za}
Shude Mao.
\newblock {Astrophysical Applications of Gravitational Microlensing}.
\newblock {\em Res. Astron. Astrophys.}, 12:947--972, 2012.

\bibitem{Koopmans:2009av}
L.~V.~E. Koopmans, A.~Bolton, T.~Treu, O.~Czoske, M.~Auger, M.~Barnabe,
  S.~Vegetti, R.~Gavazzi, L.~Moustakas, and S.~Burles.
\newblock {The Structure \textbackslash{} and Dynamics of Massive Early-type
  Galaxies: On Homology, Isothermality and Isotropy inside one Effective
  Radius}.
\newblock {\em Astrophys. J. Lett.}, 703:L51--L54, 2009.

\bibitem{Clowe:2006eq}
Douglas Clowe, Marusa Bradac, Anthony~H. Gonzalez, Maxim Markevitch, Scott~W.
  Randall, Christine Jones, and Dennis Zaritsky.
\newblock {A direct empirical proof of the existence of dark matter}.
\newblock {\em Astrophys. J. Lett.}, 648:L109--L113, 2006.

\bibitem{Clegg:1997ya}
Andrew~W. Clegg, Alan~L. Fey, and T.~Joseph~W. Lazio.
\newblock {The Gaussian plasma lens in astrophysics. refraction}.
\newblock {\em Astrophys. J.}, 496:253, 1998.

\bibitem{Bisnovatyi-Kogan:2010flt}
G.~S. Bisnovatyi-Kogan and O.~Yu. Tsupko.
\newblock {Gravitational lensing in a non-uniform plasma}.
\newblock {\em Mon. Not. Roy. Astron. Soc.}, 404:1790--1800, 2010.

\bibitem{Crisnejo:2018uyn}
Gabriel Crisnejo and Emanuel Gallo.
\newblock {Weak lensing in a plasma medium and gravitational deflection of
  massive particles using the Gauss-Bonnet theorem. A unified treatment}.
\newblock {\em Phys. Rev. D}, 97(12):124016, 2018.

\bibitem{Crisnejo:2018ppm}
Gabriel Crisnejo, Emanuel Gallo, and Adam Rogers.
\newblock {Finite distance corrections to the light deflection in a
  gravitational field with a plasma medium}.
\newblock {\em Phys. Rev. D}, 99(12):124001, 2019.

\bibitem{Crisnejo:2019xtp}
Gabriel Crisnejo, Emanuel Gallo, and Jos\'e~R. Villanueva.
\newblock {Gravitational lensing in dispersive media and deflection angle of
  charged massive particles in terms of curvature scalars and energy-momentum
  tensor}.
\newblock {\em Phys. Rev. D}, 100(4):044006, 2019.

\bibitem{Crisnejo:2019ril}
Gabriel Crisnejo, Emanuel Gallo, and Kimet Jusufi.
\newblock {Higher order corrections to deflection angle of massive particles
  and light rays in plasma media for stationary spacetimes using the
  Gauss-Bonnet theorem}.
\newblock {\em Phys. Rev. D}, 100(10):104045, 2019.

\bibitem{Tuntsov_2016ApJ}
Artem~V. {Tuntsov}, Mark~A. {Walker}, Leon V.~E. {Koopmans}, Keith~W.
  {Bannister}, Jamie {Stevens}, Simon {Johnston}, Cormac {Reynolds}, and
  Hayley~E. {Bignall}.
\newblock {Dynamic Spectral Mapping of Interstellar Plasma Lenses}.
\newblock {\em \apj}, 817(2):176, February 2016.

\bibitem{Bannister_2016Sci}
Keith~W. {Bannister}, Jamie {Stevens}, Artem~V. {Tuntsov}, Mark~A. {Walker},
  Simon {Johnston}, Cormac {Reynolds}, and Hayley {Bignall}.
\newblock {Real-time detection of an extreme scattering event: Constraints on
  Galactic plasma lenses}.
\newblock {\em Science}, 351(6271):354--356, January 2016.

\bibitem{Stanimirovi_2018}
Sne{\v{z}}ana Stanimirovi{\'c} and Ellen~G Zweibel.
\newblock Atomic and ionized microstructures in the diffuse interstellar
  medium.
\newblock {\em Annual Review of Astronomy and Astrophysics}, 56:489--540, 2018.

\bibitem{Suresh_2019ApJ}
A.~{Suresh} and J.~M. {Cordes}.
\newblock {Induced Polarization from Birefringent Pulse Splitting in
  Magneto-ionic Media}.
\newblock {\em \apj}, 870(1):29, January 2019.

\bibitem{Rogers_2020MNRAS}
Adam {Rogers}, Abdul {Mohamed}, Bailey {Preston}, Jason~D. {Fiege}, and
  Xinzhong {Er}.
\newblock {Magnetized filament models for diverging plasma lenses}.
\newblock {\em Month.Not.Roy.Astro.Soc}, 493(2):1736--1752, April 2020.

\bibitem{Pen:2013njl}
Ue-Li Pen and Yuri Levin.
\newblock {Pulsar scintillations from corrugated reconnection sheets in the
  interstellar medium}.
\newblock {\em Mon. Not. Roy. Astron. Soc.}, 442(4):3338--3346, 2014.

\bibitem{Simard_2018MNRAS}
Dana {Simard} and Ue-Li {Pen}.
\newblock {Predicting pulsar scintillation from refractive plasma sheets}.
\newblock {\em Month.Not.Roy.Astro.Soc}, 478(1):983--994, July 2018.

\bibitem{Pen:2011dz}
Ue-Li Pen and Lindsay King.
\newblock {Refractive Convergent Plasma Lenses explain ESE and pulsar
  scintillation}.
\newblock {\em Mon. Not. Roy. Astron. Soc.}, 421:L132, 2012.

\bibitem{Main_2018Natur}
Robert {Main}, I.~Sheng {Yang}, Victor {Chan}, Dongzi {Li}, Fang~Xi {Lin},
  Nikhil {Mahajan}, Ue-Li {Pen}, Keith {Vanderlinde}, and Marten~H. {van
  Kerkwijk}.
\newblock {Pulsar emission amplified and resolved by plasma lensing in an
  eclipsing binary}.
\newblock {\em \nat}, 557(7706):522--525, May 2018.

\bibitem{Li_2019MNRAS}
Dongzi {Li}, Fang~Xi {Lin}, Robert {Main}, Ue-Li {Pen}, Marten~H. {van
  Kerkwijk}, and I.~Sheng {Yang}.
\newblock {Constraining magnetic fields through plasma lensing: application to
  the Black Widow pulsar}.
\newblock {\em Month.Not.Roy.Astro.Soc}, 484(4):5723--5733, April 2019.

\bibitem{2020ApJ...889..158E}
Xinzhong {Er}, Yuan-Pei {Yang}, and Adam {Rogers}.
\newblock {The Effects of Plasma Lensing on the Inferred Dispersion Measures of
  Fast Radiobursts}.
\newblock {\em Astrophysical journal}, 889(2):158, February 2020.

\bibitem{Er:2022lad}
Xinzhong Er and Shude Mao.
\newblock {The effects of plasma on the magnification and time delay of
  strongly lensed fast radio bursts}.
\newblock {\em Mon. Not. Roy. Astron. Soc.}, 516(2):2218--2222, 2022.

\bibitem{Rogers_2015}
Adam Rogers.
\newblock Frequency-dependent effects of gravitational lensing within plasma.
\newblock {\em Monthly Notices of the Royal Astronomical Society},
  451(1):17–25, May 2015.

\bibitem{Rogers:2016xcc}
Adam Rogers.
\newblock {Escape and Trapping of Low-Frequency Gravitationally Lensed Rays by
  Compact Objects within Plasma}.
\newblock {\em Mon. Not. Roy. Astron. Soc.}, 465(2):2151--2159, 2017.

\bibitem{Battye:2021xvt}
R.~A. Battye, B.~Garbrecht, J.~I. McDonald, and S.~Srinivasan.
\newblock {Radio line properties of axion dark matter conversion in neutron
  stars}.
\newblock {\em JHEP}, 09:105, 2021.

\bibitem{Briozzo:2022yzi}
Gast\'on Briozzo and Emanuel Gallo.
\newblock {Analytical expressions for pulse profile of neutron stars in plasma
  environments}.
\newblock {\em Eur. Phys. J. C}, 83(2):165, 2023.

\bibitem{Perlick:2017fio}
Volker Perlick and Oleg~Yu. Tsupko.
\newblock {Light propagation in a plasma on Kerr spacetime: Separation of the
  Hamilton-Jacobi equation and calculation of the shadow}.
\newblock {\em Phys. Rev. D}, 95(10):104003, 2017.

\bibitem{Perlick:2021aok}
Volker Perlick and Oleg~Yu. Tsupko.
\newblock {Calculating black hole shadows: Review of analytical studies}.
\newblock {\em Phys. Rept.}, 947:1--39, 2022.

\bibitem{Kimpson:2019mji}
Tom Kimpson, Kinwah Wu, and Silvia Zane.
\newblock {Spatial dispersion of light rays propagating through a plasma in
  Kerr space\textendash{}time}.
\newblock {\em Mon. Not. Roy. Astron. Soc.}, 484(2):2411--2419, 2019.

\bibitem{Huang:2018rfn}
Yang Huang, Yi-Ping Dong, and Dao-Jun Liu.
\newblock {Revisiting the shadow of a black hole in the presence of a plasma}.
\newblock {\em Int. J. Mod. Phys. D}, 27(12):1850114, 2018.

\bibitem{Zhang:2022osx}
Zhenyu Zhang, Haopeng Yan, Minyong Guo, and Bin Chen.
\newblock {Shadow of Kerr black hole surrounded by an angular Gaussian
  distributed plasma}.
\newblock {\em arXiv:2206.04430}, 6 2022.

\bibitem{Badia:2021kpk}
Javier Bad\'\i{}a and Ernesto~F. Eiroa.
\newblock {Shadow of axisymmetric, stationary, and asymptotically flat black
  holes in the presence of plasma}.
\newblock {\em Phys. Rev. D}, 104(8):084055, 2021.

\bibitem{Briozzo:2022mgg}
Gast\'on Briozzo, Emanuel Gallo, and Thomas M\"adler.
\newblock {Shadows of rotating black holes in plasma environments with
  aberration effects. arxiv:2211.05620}.
\newblock 11 2022.

\bibitem{Er:2013efa}
Xinzhong Er and Shude Mao.
\newblock {Effects of plasma on gravitational lensing}.
\newblock {\em Mon. Not. Roy. Astron. Soc.}, 437(3):2180--2186, 2014.

\bibitem{Tsupko:2019axo}
Oleg~Yu. Tsupko and Gennady~S. Bisnovatyi-Kogan.
\newblock {Hills and holes in the microlensing light curve due to plasma
  environment around gravitational lens}.
\newblock {\em Mon. Not. Roy. Astron. Soc.}, 491(4):5636--5649, 2020.

\bibitem{Sun:2022ujt}
Jiarui Sun, Xinzhong Er, and Oleg~Yu. Tsupko.
\newblock {Binary microlensing with plasma environment \textendash{} star and
  planet}.
\newblock {\em Mon. Not. Roy. Astron. Soc.}, 520(1):994--1004, 2023.

\bibitem{Alard:2007ya}
C.~Alard.
\newblock {Gravitational arcs as a perturbation of the perfect ring}.
\newblock {\em Mon. Not. Roy. Astron. Soc.}, 382:58, 2007.

\bibitem{2020A&A...641A...6P}
{Planck Collaboration}.
\newblock {Planck 2018 results. VI. Cosmological parameters}.
\newblock {\em Astronomy and Astrophysics}, 641:A6, September 2020.

\bibitem{Beck:2013bxa}
Rainer Beck and Richard Wielebinski.
\newblock {\em {Magnetic Fields in the Milky Way and in Galaxies}}.
\newblock 2 2013.

\bibitem{2020MNRAS.493.1736R}
Adam {Rogers}, Abdul {Mohamed}, Bailey {Preston}, Jason~D. {Fiege}, and
  Xinzhong {Er}.
\newblock {Magnetized filament models for diverging plasma lenses}.
\newblock {\em Month.Not.Roy.Soc}, 493(2):1736--1752, April 2020.

\bibitem{Li:2018ssw}
Dongzi Li, Fang~Xi Lin, Robert Main, Ue-Li Pen, Marten~H. van Kerkwijk, and
  I-Sheng Yang.
\newblock {Constraining small scale magnetic fields through plasma lensing:
  Application to the Black widow eclipsing pulsar binary}.
\newblock 9 2018.

\bibitem{bittencourt2013fundamentals}
J.A. Bittencourt.
\newblock {\em Fundamentals of Plasma Physics}.
\newblock Springer New York, 2013.

\bibitem{draine2010physics}
B.T. Draine.
\newblock {\em Physics of the Interstellar and Intergalactic Medium}.
\newblock Princeton Series in Astrophysics. Princeton University Press, 2010.

\bibitem{Donner:2019hlp}
J.~Y. Donner et~al.
\newblock {First detection of frequency-dependent, time-variable dispersion
  measures}.
\newblock {\em Astron. Astrophys.}, 624:A22, 2019.

\bibitem{Habara11}
Yuta Habara and Kazuhiro Yamamoto.
\newblock {Analytic approach to perturbed Einstein ring with elliptical NFW
  lens model}.
\newblock {\em Int. J. Mod. Phys. D}, 20:371--400, 2011.

\bibitem{gutierrez2010galaxy}
Leonel Guti{\'e}rrez and John~E Beckman.
\newblock The galaxy-wide distributions of mean electron density in the h ii
  regions of m51 and ngc 4449.
\newblock {\em The Astrophysical Journal Letters}, 710(1):L44, 2010.

\bibitem{Davies_2021}
Rebecca~L. Davies, N.~M.~Förster Schreiber, R.~Genzel, T.~T. Shimizu, R.~I.
  Davies, A.~Schruba, L.~J. Tacconi, H.~Übler, E.~Wisnioski, S.~Wuyts,
  M.~Fossati, R.~Herrera-Camus, D.~Lutz, J.~T. Mendel, T.~Naab, S.~H. Price,
  A.~Renzini, D.~Wilman, A.~Beifiori, S.~Belli, A.~Burkert, J.~Chan,
  A.~Contursi, M.~Fabricius, M.~M. Lee, R.~P. Saglia, and A.~Sternberg.
\newblock The {KMOS}3d survey: Investigating the origin of the elevated
  electron densities in star-forming galaxies at 1 $\lesssim$ z $\lesssim$ 3.
\newblock {\em The Astrophysical Journal}, 909(1):78, mar 2021.

\bibitem{Chy_y_2018}
K.~T. Chy{\.{z} }y, W.~Jurusik, J.~Piotrowska, B.~Nikiel-Wroczy{\'{n}}ski,
  V.~Heesen, V.~Vacca, N.~Nowak, R.~Paladino, P.~Surma, S.~S. Sridhar,
  G.~Heald, R.~Beck, J.~Conway, K.~Sendlinger, M.~Cury{\l}o, D.~Mulcahy, J.~W.
  Broderick, M.~J. Hardcastle, J.~R. Callingham, G.~Gürkan, M.~Iacobelli,
  H.~J.~A. Röttgering, B.~Adebahr, A.~Shulevski, R.-J. Dettmar, R.~P. Breton,
  A.~O. Clarke, J.~S. Farnes, E.~Orr{\'{u}}, V.~N. Pandey, M.~Pandey-Pommier,
  R.~Pizzo, C.~J. Riseley, A.~Rowlinson, A.~M.~M. Scaife, A.~J. Stewart, A.~J.
  van~der Horst, and R.~J. van Weeren.
\newblock {LOFAR} {MSSS}: Flattening low-frequency radio continuum spectra of
  nearby galaxies.
\newblock {\em Astronomy and Astrophysics}, 619:A36, nov 2018.

\bibitem{Cordes:2002wz}
James~M. Cordes and T.~J.~W. Lazio.
\newblock {NE2001. 1. A New model for the galactic distribution of free
  electrons and its fluctuations}.
\newblock 7 2002.

\bibitem{Yao_2017}
J.~M. Yao, R.~N. Manchester, and N.~Wang.
\newblock A {NEW} {ELECTRON}-{DENSITY} {MODEL} {FOR} {ESTIMATION} {OF} {PULSAR}
  {AND} {FRB} {DISTANCES}.
\newblock {\em The Astrophysical Journal}, 835(1):29, jan 2017.

\bibitem{Yamasaki:2019htx}
Shotaro Yamasaki and Tomonori Totani.
\newblock {The Galactic Halo Contribution to the Dispersion Measure of
  Extragalactic Fast Radio Bursts}.
\newblock 9 2019.

\bibitem{Price:2021gzo}
Danny~C. Price, Adam Deller, and Chris Flynn.
\newblock {A comparison of Galactic electron density models using PyGEDM}.
\newblock 6 2021.

\bibitem{Gupta:1999nf}
Yashwant Gupta, N.~D.~Ramesh Bhat, and A.~Pramesh Rao.
\newblock {Multiple imaging of psr b1133+16 by the ism}.
\newblock {\em THE ASTROPHYSICAL JOURNAL}, 520:173--181, 1999.

\bibitem{Pushkarev:2013zqa}
A.~B. Pushkarev et~al.
\newblock {VLBA observations of a rare multiple quasar imaging event caused by
  refraction in the interstellar medium}.
\newblock {\em Astronomy and Astrophysics}, 555:A80, 2013.

\bibitem{Wang:2022kkk}
Yu-Bin Wang, Zhi-Gang Wen, Rai Yuen, N.~Wang, Jian-Ping Yuan, and Xia Zhou.
\newblock {The Multiple Images of the Plasma Lensing FRB}.
\newblock {\em Res. Astron. Astrophys.}, 22(6):065017, 2022.

\bibitem{refId0}
{Jackson, N.}, {Badole, S.}, {Morgan, J.}, {Chhetri, R.}, {Prusis, K.},
  {Nikolajevs, A.}, {Morabito, L.}, {Brentjens, M.}, {Sweijen, F.}, {Iacobelli,
  M.}, {Orr\`u, E.}, {Sluman, J.}, {Blaauw, R.}, {Mulder, H.}, {van Dijk, P.},
  {Mooney, S.}, {Deller, A.}, {Moldon, J.}, {Callingham, J. R.}, {Harwood, J.},
  {Hardcastle, M.}, {Heald, G.}, {Drabent, A.}, {McKean, J. P.}, {Asgekar, A.},
  {Avruch, I. M.}, {Bentum, M. J.}, {Bonafede, A.}, {Brouw, W. N.}, {Br\"uggen,
  M.}, {Butcher, H. R.}, {Ciardi, B.}, {Coolen, A.}, {Corstanje, A.}, {Damstra,
  S.}, {Duscha, S.}, {Eisl\"offel, J.}, {Falcke, H.}, {Garrett, M.}, {de
  Gasperin, F.}, {Griessmeier, J.-M.}, {Gunst, A. W.}, {van Haarlem, M. P.},
  {Hoeft, M.}, {van der Horst, A. J.}, {J\"utte, E.}, {Koopmans, L. V. E.},
  {Krankowski, A.}, {Maat, P.}, {Mann, G.}, {Miley, G. K.}, {Nelles, A.},
  {Norden, M.}, {Paas, M.}, {Pandey, V. N.}, {Pandey-Pommier, M.}, {Pizzo, R.
  F.}, {Reich, W.}, {Rothkaehl, H.}, {Rowlinson, A.}, {Ruiter, M.}, {Shulevski,
  A.}, {Schwarz, D. J.}, {Smirnov, O.}, {Tagger, M.}, {Vocks, C.}, {van Weeren,
  R. J.}, {Wijers, R.}, {Wucknitz, O.}, {Zarka, P.}, {Zensus, J. A.}, and
  {Zucca, P.}
\newblock Sub-arcsecond imaging with the international lofar telescope - ii.
  completion of the lofar long-baseline calibrator survey.
\newblock {\em Astronomy and Astrophysics}, 658:A2, 2022.

\bibitem{BIRRER2018189}
Simon Birrer and Adam Amara.
\newblock Lenstronomy: Multi-purpose gravitational lens modelling software
  package.
\newblock {\em Physics of the Dark Universe}, 22:189--201, 2018.

\bibitem{lima}
{Lima-Costa, F.}, {Martins, L. P.}, {Rodr\'{\i}guez-Ardila, A.}, and {Fraga,
  L.}
\newblock Spectroscopic study of the hii regions in the ngc 1232 galaxy.
\newblock {\em Astronomy and Astrophysics}, 642:A203, 2020.

\bibitem{Tomas-tesina}
Tom\'as~Andr\'es Ulla.
\newblock {Efectos de entornos plasm\'aticos en el r\'egimen de lente fuerte}.
\newblock {\url{http://hdl.handle.net/11086/25964}}, 2022.

\bibitem{Tomas-work}
Tom\'as~Andr\'es Ulla and Emanuel Gallo.
\newblock {Effects on lensed jets by plasma strong lensing}.
\newblock {Work in progress}, 2023.

\bibitem{Dumet-Montoya:2013yqp}
Habib~S. Dumet-Montoya, Gabriel~B. Caminha, Bruno Moraes, Martin Makler,
  Mandeep S.~S. Gill, and Basilio~X. Santiago.
\newblock {On the validity of the Perturbative Approach for Strong Lensing:
  Local Distortion for Pseudo-Elliptical Models}.
\newblock {\em Mon. Not. Roy. Astron. Soc.}, 433:2975, 2013.

\bibitem{Er:2019jkg}
Xinzhong Er and Adam Rogers.
\newblock {Two families of elliptical plasma lenses}.
\newblock {\em Mon. Not. Roy. Astron. Soc.}, 488(4):5651--5664, 2019.

\bibitem{Er:2017lue}
Xinzhong Er and Adam Rogers.
\newblock {Two families of astrophysical diverging lens models}.
\newblock {\em Mon. Not. Roy. Astron. Soc.}, 475(1):867--878, 2017.

\bibitem{Schneider-book}
Peter {Schneider}, J{\"u}rgen {Ehlers}, and Emilio~E. {Falco}.
\newblock {\em {Gravitational Lenses}}.
\newblock 1992.

\bibitem{Binney2008}
J.~{Binney} and S.~{Tremaine}.
\newblock {\em {Galactic Dynamics: Second Edition}}.
\newblock Princeton University Press, 2008.

\bibitem{moon1988field}
P.H. Moon and D.E. Spencer.
\newblock {\em Field theory handbook: including coordinate systems,
  differential equations, and their solutions}.
\newblock Springer-Verlag, 1988.

\bibitem{Gallo11}
Emanuel Gallo and Osvaldo~M. Moreschi.
\newblock {Gravitational lens optical scalars in terms of energy-momentum
  distributions}.
\newblock {\em Phys.Rev.}, D83:083007, 2011.

\bibitem{Boero:2018a}
Ezequiel~F. Boero and Osvaldo~M. Moreschi.
\newblock Gravitational lens optical scalars in terms of energy–momentum
  distributions in the cosmological framework.
\newblock {\em MNRAS}, 475(4):4683--4703, 2018.

\bibitem{Geroch:1973am}
Robert~P. Geroch, A.~Held, and R.~Penrose.
\newblock {A space-time calculus based on pairs of null directions}.
\newblock {\em J.Math.Phys.}, 14:874--881, 1973.

\bibitem{xact}
José~M. Martín-García.
\newblock {xAct suite of packages for tensor computer algebra}.
\newblock {Version 1.2.0 of xAct release \url{http://http://www.xact.es}}, Oct
  2021.
\newblock Efficient tensor computer algebra for the Wolfram Language.

\bibitem{rksuite-90}
R.W. Brankin, I.~Gladwell, and L.F. Shampine.
\newblock {RKSUITE: a suite of Runge-Kutta codes for the initial value problem
  for ODEs}.
\newblock {rksuite\_90 v1.2}, Dec 1995.
\newblock Softreport 92-S1, Department of Mathematics, Southern Methodist
  University, Dallas, Texas, U.S.A.

\bibitem{van_Haarlem_2013}
Michael~P van Haarlem, Michael~W Wise, AW~Gunst, George Heald, John~P McKean,
  Jason~WT Hessels, A~Ger de~Bruyn, Ronald Nijboer, John Swinbank, Richard
  Fallows, et~al.
\newblock Lofar: The low-frequency array.
\newblock {\em Astronomy and Astrophysics}, 556:A2, 2013.

\bibitem{Nenufar}
NenuFAR.
\newblock {NenuFAR Observatory}.
\newblock {NenuFAR observatory
  \url{https://nenufar.obs-nancay.fr/en/homepage-en/}}, 2022.

\bibitem{2022A&A...658A...9G}
C.~{Groeneveld}, R.~J. {van Weeren}, G.~K. {Miley}, L.~K. {Morabito}, F.~{de
  Gasperin}, J.~R. {Callingham}, F.~{Sweijen}, M.~{Br{\"u}ggen}, A.~{Botteon},
  A.~{Offringa}, G.~{Brunetti}, J.~{Moldon}, M.~{Bondi}, A.~{Kappes}, and
  H.~J.~A. {R{\"o}ttgering}.
\newblock {Pushing sub-arcsecond resolution imaging down to 30 MHz with the
  trans-European International LOFAR Telescope}.
\newblock {\em Astronomy and Astrophysics}, 658:A9, February 2022.

\bibitem{Bisnovatyi-Kogan:2022yzj}
Gennady~S. Bisnovatyi-Kogan and Oleg~Yu. Tsupko.
\newblock {Time delay induced by plasma in strong lens systems}.
\newblock 12 2022, arxiv:2301.00053.

\bibitem{meneghetti2021introduction}
M.~Meneghetti.
\newblock {\em Introduction to Gravitational Lensing: With Python Examples}.
\newblock Lecture Notes in Physics. Springer International Publishing, 2021.

\bibitem{Wagner:2020ihx}
Jenny Wagner and Xinzhong Er.
\newblock {Plasma lensing in comparison to gravitational lensing -- Formalism
  and degeneracies}.
\newblock 6 2020.

\end{thebibliography}

\end{document}